\documentclass[12pt]{article}
\usepackage{amsmath}
\usepackage{amssymb}
\begin{document}
\begin{center}
{\bf {\large{Superfield Approaches to a Model of \\ Bosonic String: Curci-Ferrari Type Restrictions}}}

\vskip 2.5cm

{\sf  A. Tripathi$^{(a)}$, A. K. Rao$^{(a)}$, R. P. Malik$^{(a,b)}$}\\
$^{(a)}$ {\it Physics Department, Institute of Science,}\\
{\it Banaras Hindu University, Varanasi - 221 005, (U.P.), India}\\

\vskip 0.1cm

$^{(b)}$ {\it DST Centre for Interdisciplinary Mathematical Sciences,}\\
{\it Institute of Science, Banaras Hindu University, Varanasi - 221 005, India}\\
{\small {\sf {e-mails:  ankur1793@gmail.com; amit.akrao@gmail.com; 
  rpmalik1995@gmail.com}}}
\end{center}

\vskip 1.5 cm

\noindent
{\bf Abstract:}
Exploiting the theoretical potential of the {\it modified} Bonora-Tonin superfield approach (MBTSA) as well as the (anti-)chiral superfield
approach (ACSA) to Becchi-Rouet-Stora-Tyutin (BRST) formalism, we derive the {\it complete} set of off-shell 
nilpotent (anti-) BRST symmetry transformations corresponding to the
{\it classical} two $(1 + 1)$-dimensional (2D) diffeomorphism symmetry transformations on the world-sheet (that is traced out by the motion of 
a model of bosonic string). Only the BRST symmetry transformations for {\it this} model have been discussed 
in the {\it earlier} literature. We derive the (anti-)BRST invariant Curci-Ferrari (CF) type restrictions (using MBTSA)
which turn out to be the root-cause behind the absolute anticommutativity of the above (anti-)BRST symmetry transformations. We capture the symmetry
invariance of the (anti-)BRST invariant Lagrangian densities within the ambit of ACSA. The derivation of the {\it proper} anti-BRST transformations
(corresponding to the {\it already} known BRST transformations)
{\it and} the (anti-)BRST invariant CF-type restrictions are the {\it novel} results in our present endeavor.



\newpage

\section {Introduction}

\noindent
Superfield approaches (see, e.g. [1-8]) to Becchi-Rouet-Stora-Tyutin (BRST) formalism are geometrically elegant,
mathematically rich and physically very intuitive as they provide the geometrical basis for the off-shell nilpotency
and absolute anticommutativity of the {\it quantum} (anti-)BRST symmetry transformations that are associated with
a given {\it classical} local gauge symmetry transformation for a {\it classically} gauge invariant theory.
In the above {\it usual} superfield approaches [1-8], {\it only} the $p$-form ($p = 1, 2, 3...$) gauge theories have been 
considered which are characterized by the existence of the first-class constraints on {\it them} in the terminology of Dirac's
prescription for the classification scheme of constraints (see, e.g. [9, 10]). It has been a challenging problem to incorporate the diffeomorphism
invariant theories in the domain of the superfield approaches to BRST formalism. An attempt 
has been made by Delbourgo, {\it etal}., (see, e.g. [11]) in this direction where a diffeomorphism invariant gravitational theory has been considered.
However, in our present endeavor, we shall {\it not} discuss {\it anything} connected with the superfield approach developed in [11] 
for the BRST-analysis of our present two-dimensional (2D) diffeomorphism invariant theory.

A very successful application
of the superfield approach [4-6] to BRST formalism (in the context of D-dimensional non-Abelian 1-form gauge theory) has been performed by Bonora and Tonin
(BT). We have exploited the theoretical techniques and tricks of {\it this} approach in the context of BRST-analysis of
the higher $p$-form ($p = 2, 3$) Abelian gauge theories
[12]. It has been a very exciting problem to incorporate the diffeomorphism symmetry transformations within the framework of
BT-superfield formalism. A breakthrough, in this direction, has been made by Bonora in a very recent paper [13] where the D-dimensional diffeomorphism
invariant theory has been discussed within the ambit of BT-superfield approach [4-6]. We have christened {\it this} theoretical technique as the {\it modified}
version of BT-superfield approach (MBTSA) to BRST formalism [13] and   
applied {\it its} theoretical potential in the context of the 1D diffeomorphism (i.e. reparameterization) invariant model of
 a free spinning  supersymmetric  (SUSY) relativistic particle [14] and established that its Curci-Ferrari (CF) type of restriction
as well as the gauge-fixing {\it and} Faddeev-Popov ghost terms are the {\it same} as for the other 1D 
diffeomorphism (i.e. reparameterization) invariant models of a free scalar and non-SUSY relativistic particle as well as a non-SUSY and 
 non-relativistic free particle (see, e.g. [14] and references therein).

In the applications of MBTSA [13], it turns out that we have to take into account the {\it full} super expansions of the superfields {\it defined} on the 
(D, 2)-dimensional supermanifold. In other words, we perform the super expansion of the {\it above} 
superfields along {\it all} the possible Grassmannian directions of the (D, 2)-dimensional
supermanifold on which a D-dimensional {\it ordinary} diffeomorphism invariant theory is generalized. The idea of horizontality condition (HC)
enables us to derive the (anti-)BRST symmetry transformations for the scalars, vectors, tensors, etc. However, we invoke the Nakanishi-Lautrup 
type auxiliary fields $(\bar B_\mu)B_\mu$ (with $\mu = 0,\,1,\,2...D-1$) in the {\it standard} nilpotent (anti-)BRST symmetry transformations:
$s_b\,\bar C_\mu = i\, B_\mu,\, s_b\,B_\mu = 0,\, s_{ab}\,C_\mu = i\, \bar B_\mu,\, s_{ab}\,\bar B_\mu = 0$ of the (anti-)ghost fields $(\bar C_\mu)C\mu$
in the case of the D-dimensional diffeomorphism invariant theory in an ad-hoc manner.
 This forces us to consider the (anti-)chiral super expansions of the superfields 
[cf. Eq. (27) below]. At this juncture, it becomes essential for us to take into account the theoretical 
tricks and techniques of the (anti-)chiral superfield approach (ACSA) to BRST formalism
(see e.g. [15] and references therein) which has been developed by us.

The central theme of our present investigation is to apply the ideas of MBTSA and ACSA to BRST formalism in the realm of a 2D diffeomorphism invariant
theory of a model of bosonic string and derive (i) {\it all} the (anti-)BRST symmetries of this theory in a consistent and clear fashion, and 
(ii) the CF-type restrictions which are responsible for the absolute anticommutativity of the (anti-)BRST
symmetry transformations. We have also derived the BRST and anti-BRST invariant Lagrangian densities and captured their symmetry invariance(s) in the 
language of ACSA to BRST formalism. We would like to lay emphasis on the fact that the theoretical potential of MBTSA has been 
responsible for the derivation of (i) the (anti-)BRST symmetry transformations for the {\it pure} Lorentz scalars, and (ii) the (anti-)BRST
invariant CF-type restrictions. However, we have been able to derive {\it all} the proper (anti-)BRST transformations for 
{\it all} the {\it other} fields by using ACSA.

The following motivating factors have been at the heart of our present investigation. First, we have already used the beautiful blend of 
theoretical ideas behind MBTSA and ACSA in the cases of some 1D diffeomorphism (i.e. reparameterization) invariant theories of SUSY (i.e. spinning)
relativistic particle, NSUSY (i.e. scalar) relativistic particle and NSUSY and non-relativistic system of a free particle for the discussion 
of BRST analysis. However, these models are {\it also} endowed with the {\it gauge} symmetry transformations which are a 
kind of subset of the reparameterization symmetry
transformations (under {\it specific} limits).
To be precise, it has been shown (see, e.g. [14] and references therein) that the gauge symmetry transformations (generated by the first-class constraints) are 
{\it equivalent} to the reparameterization symmetry transformations if we use $(i)$ the specific set of equations of motion, and $(ii)$
identify the transformation parameters of {\it both} these symmetries in a specific manner.
Thus, it is a challenging problem for us to use the theoretical strength of MBTSA and ACSA in the context of  a 2D
diffeomorphism invariant theory which does {\it not} respect the  gauge symmetry transformations as have been demonstrated in [14] for a 1D diffeomorphism 
invariant theory. We have discussed, in our present endeavor, a model of 
bosonic string which has the 2D diffeomorphism symmetry invariance {\it but} it does not respect a gauge 
symmetry transformation. Second,  one of the sacrosanct aspects
of BRST formalism is the existence of the {\it quantum} BRST and anti-BRST symmetries {\it together} for a given {\it classical} gauge/diffeomorphism 
symmetry transformation. For our present bosonic string, {\it only} the BRST symmetries are known in literature [16]. Thus, it is a challenge for us to
derive the {\it proper} anti-BRST symmetry transformations corresponding to the {\it above} BRST symmetry transformations. 
We have accomplished this goal in our present endeavor. Finally, the hallmark of a BRST-quantized
theory is the existence of the CF-type restrictions 
 which provide the independent {\it identity} to the BRST and anti-BRST symmetries (and corresponding charges) at the 
{\it quantum} level. We have derived these restrictions, too.

The theoretical contents of our present endeavor are organized as follows. In Sec. 2, we concisely discuss the (anti-)BRST
symmetry transformations for the gauge-fixed Lagrangian densities of the bosonic string theory. Our Sec. 3 is devoted to 
the derivation of the Curci-Ferrari (CF) type restrictions for our BRST invariant theory within the framework of MBTSA. In addition,
we {\it also} derive the (anti-)BRST symmetry transformations for the target space coordinates and the 
determinant of the modified version of the inverse of the 2D metric 
tensor. Sec. 4 contains the derivation of the nilpotent (anti-)BRST symmetries for the {\it other} fields of our theory by 
exploiting the theoretical potential of ACSA. We capture the (anti-)BRST invariances of the Lagrangian 
densities using ACSA in Sec. 5. Finally, we make some concluding remarks in our Sec. 6.

\section {Preliminary: (Anti-)BRST Symmetries}

We begin with the following (anti-)BRST invariant Lagrangian densities [${\cal L}_{(a)b}$] for the model of the bosonic string of our theory 
(see, e.g. [17] for details)
\begin{eqnarray}
{\cal L}_{ab} & = &{\cal L}_0 - \bar B_1 A_1  - {\bar B}_0 A_0
+ i\,\Bigl [ C_1 \, (\partial_0 \, \bar C^1 + \partial_1 \, \bar C^0) + C_0 (\partial_a \, \bar C^a) + (\partial_a C_0)\, \bar C^a 
\Bigr] \, A_0 \nonumber\\
&+& i\,\Bigl [ C_0 \, (\partial_0 \, \bar C^1 + \partial_1 \, \bar C^0) + C_1 \, (\partial_a \, \bar C^a) + (\partial_a \, C_1) \,\bar C^a 
\Bigr] \, A_1  \nonumber\\
& + & i\,\Bigl[ C_1 \,(\partial_0 \, \bar C^0 - \partial_1 \, \bar C^1) + C_0 \, (\partial_0 \, \bar C^1 - \partial_1 \, \bar C^0) 
 \Bigr] \, A_2,
\end{eqnarray}
\begin{eqnarray}
{\cal L}_b &=& {\cal L}_0 + B_1 A_1  + B_0 A_0 - i\, \big[\bar C_1 (\partial_0 C^1 + \partial_1 C^0)
+ \bar C_0 (\partial_a C^a) - C^a\partial_a \bar C_0 
\big]\, A_0 \nonumber\\
&-& i \,\big[ \bar C_0 (\partial_0 C^1 + \partial_1 C^0) - C^a \,\partial_a \,\bar C_1 + \bar C_1 \,(\partial_a \, C^a) \big] \,A_1 \nonumber\\ 
&-& i \,\big[ \bar C_1\, (\partial_0 C^0\, - \partial_1 C^1) + \bar C_0 \,(\partial_0 \, C^1\, - \partial_1 \, C^0) \big]\, A_2. 
\end{eqnarray}
where the 2D diffeomorphism invariant {\it classical} action integral $(S_0)$ w.r.t. ${\cal L}_0$ is [16]:
\begin{eqnarray}
S_0 = \int d^2\,\xi\,{\cal L}_0 \equiv \int_{- \infty}^{+ \infty} d\,\tau\, \int_{\sigma = 0}^{\sigma = \pi} d\,\sigma\,\Big[- \frac{1}{2\kappa}\, 
{\tilde g}^{mn}\,\partial_m \,X^\mu \, \partial_n \, X_\mu + E\, \big(\mbox{det} \,\tilde g + 1\big)\Big].
\end{eqnarray}
In the above, we have taken the notation $\xi^a = (\xi^0, \, \xi^1) = (\tau, \, \sigma )$ where $\tau$ is the evolution parameter (with $-\infty <
\tau < +\infty$)  and $\sigma$ denotes the length of the bosonic string (with $0 \leq \sigma \leq \pi$). 
The modified version of the {\it inverse} of the 2D metric tensor is: ${\tilde g}^{mn} = {\sqrt {-g}} \,g^{mn}$ where $g^{mn}$ is the inverse of the 
2D metric tensor $g_{mn}$ and $g = \,\,$det $(g_{mn})$. The coordinates 
$X^\mu (\xi ) \equiv X^\mu (\tau ,\sigma )$ (where $\mu = 0, 1, 2...D-1$)
correspond to the D-dimensional {\it flat} Minkowskian {\it target} space and $a, b, c, ... l, m, n = 0, 1$ are the ``time" and space directions 
on the world-sheet. The symbol $\kappa$
denotes the string tension parameter and $E$ is the Lagrange multiplier density which ensures that det $\tilde g = - 1$ so that we can have {\it two} degrees
of freedom for the metric field tensor which, being symmetric, has only {\it three} degrees of freedom on a 2D flat spacetime manifold. In other words, 
we have (see, e.g. [16] for details) the following decomposition of ${\tilde g}^{mn}$, namely;
\begin{eqnarray}
{\tilde g}^{mn} = 
\begin{pmatrix} 
A_1 + A_2  & A_0  \\ A_0 & A_1 - A_2 
\end{pmatrix}.
\end{eqnarray}
The flat limit (i.e. ${\tilde g}^{mn}\rightarrow \eta^{mn})$ can be obtained by the gauge-fixing conditions: $A_0 = A_1 = 0$. 
The {\it latter} choices imply that we have $ A_{2}^2 = 1$ when we demand $ \mbox{det} \,\tilde g = -\, 1$. This input leads to
${\tilde g}^{mn}\rightarrow \eta^{mn} = \mbox{diag}\,(+1, \, -1)$ for the choice $A_2 = +1$ where $\eta_{mn} = \eta^{mn} = \mbox{diag}\,(+ 1, - 1)$ are the
{\it flat} metric tensor $(\eta_{mn})$ and {\it its} inverse $(\eta^{mn})$ on the 2D Minkowskian spacetime manifold. 
In the derivation of the gauge-fixing and Faddeev-Popov
ghost terms, we have taken the standard prescription of the BRST formalism (see, e.g. [16]), namely;
\begin{eqnarray}
{\cal L}_{ab} &=& {\cal L}_0 + s_{ab} \,\big[\,i\,C_0 A_0 + i\,C_1 A_1 \big],\nonumber\\
{\cal L}_{b} &= &{\cal L}_0 + s_{b} \,\big[-\,i\,\bar C_0 A_0 - i\,\bar C_1 A_1 \big],
\end{eqnarray}
where the {\it full} set of nilpotent $[(s_{(a)b})^2 = 0]$ (anti-)BRST transformations $[s_{(a)b}]$ are
\begin{eqnarray}
&&s_{ab} X^\mu = \bar C^a \, \partial_a \, X^\mu, \qquad  s_{ab} C^n = i \bar B^n, \qquad  s_{ab} \bar C^n = 
\bar C^n \, \partial_m \, \bar C^n,\,\, \nonumber\\
&& s_{ab}\, E = \partial_a \,(\bar C^a E),\qquad  s_{ab} \, \bar B^n = 0,\qquad s_{ab}(\mbox{det} \, {\tilde g}) =
\bar C^m \, \partial_m \, (\mbox{det}\, {\tilde g}),\qquad \nonumber\\
&& s_{ab}\,{\tilde g}^{mn} = \partial_a (\bar C^a \, {\tilde g}^{mn}) \, - \, 
(\partial_a\, \bar C^m) \,{\tilde g}^{an} \, - \, (\partial_a \, \bar C^n)\, {\tilde g}^{ma}, \nonumber\\
&& s_{ab}\,B^n = {\bar C}^m\,\partial_m\,B^n - B^m\,\partial_m\,{\bar C}^n,
\end{eqnarray}
\begin{eqnarray}
&&s_{b} X^\mu = C^a \, \partial_a \, X^\mu, \qquad s_{b} \bar C^n = i B^n,\qquad  s_{b} \,B^n = 0, \qquad s_{b} C^n = C^b \, 
\partial_b \, C^n, \nonumber\\
&& s_{b}\,{\tilde g}^{mn} = \partial_a (C^a \, {\tilde g}^{mn}) \, - \, 
(\partial_a\,C^m) \,{\tilde g}^{an} \, - \, (\partial_a \, C^n)\, {\tilde g}^{ma}, \qquad s_{b}\, E = \partial_a \,(C^a E), \nonumber\\
&&s_{b}\,{\bar B}^n = {C}^m\,\partial_m\,{\bar B}^n 
- {\bar B}^m\,\partial_m\,{C}^n, \qquad  
s_{b}(\mbox{det} \, {\tilde g}) = C^a \, \partial_a \, (\mbox{det}\, {\tilde g}).
\end{eqnarray}
Here the fermionic $[(C^a)^2 = (\bar C^a)^2 = 0, \, C^a\,C^b + C^b\,C^a = 0, \, C^a\,\bar C^b + \bar C^b\,C^a = 0, \, 
\bar C^a\,\bar C^b + \bar C^b\,\bar C^a = 0$, etc.] (anti-)ghost fields are $(\bar C^a)C^a$
and the {\it bosonic} Nakanishi-Lautrup auxilary fields are $(\bar B^a)B^a$.
From the above, we can derive the (anti-)BRST symmetry transformations for the component gauge fields $A_0,\, A_1$ and $ A_2$ as follows:
\begin{eqnarray}
&&s_{ab} A_0 = \bar C^m\,\partial_m\, A_0  - (\partial_0 \, \bar C^1 - \partial_1 \,\bar C^0) \, A_2
 - (\partial_0 \, \bar C^1 + \partial_1 \, \bar C^0)\,  A_1, \nonumber\\
&&s_{ab} A_1 = \bar C^m\,\partial_m \, A_1  - (\partial_1 \,\bar C^0 + \partial_0 \, \bar C^1)\, A_0 - 
(\partial_0 \, \bar C^0 - \partial_1 \, \bar C^1) \, A_2,  \nonumber\\
&&s_{ab} A_2 = \bar C^m\,\partial_m \, A_2 - 
(\partial_1 \, \bar C^0 - \partial_0 \, \bar C^1) A_0 - (\partial_0 \, \bar C^0 - \partial_1 \, \bar C^1) A_1,
\end{eqnarray}
\begin{eqnarray}
&&s_{b} A_0 = C^m\,\partial_m\, A_0  - 
(\partial_0 \, C^1 - \partial_1 \,C^0) \, A_2 - (\partial_0 \, C^1 + \partial_1 \, C^0)\,  A_1, \nonumber\\
&&s_{b} A_1 = C^m\,\partial_m \, A_1  - 
(\partial_1 \,C^0 + \partial_0 \, C^1)\, A_0 - (\partial_0 \, C^0 - \partial_1 \, C^1) \, A_2,  \nonumber\\
&&s_{b} A_2 = C^m\,\partial_m \, A_2 - 
(\partial_1 \, C^0 - \partial_0 \, C^1) A_0  - (\partial_0 \, C^0 - \partial_1 \, C^1) A_1.
\end{eqnarray}
It is interesting to note that this CF-type restrictions: $ B^a + \bar B^a + i\, (C^m \, \partial_m \, \bar C^a +
 \bar C^m \, \partial_m \, C^a) = 0 $ appear in the 
following {\it simple} cases of the proof of absolute anticommutativity property:
\begin{eqnarray}
\{s_b, s_{ab}\}\,X^\mu &=&  i\,\Big[B^a + \bar B^a + i\, (C^m \, \partial_m \, \bar C^a + \bar C^m \, \partial_m \, C^a)\Big]\,\big(\partial_a\,
X^\mu\big), \nonumber\\
\{s_b, s_{ab}\}\,E &=& i\,\partial_a\,\Big[\{B^a + \bar B^a + i\, (C^m \, \partial_m \, \bar C^a + \bar C^m \, \partial_m \, C^a)\}\,E\Big], \nonumber\\
\{s_b, s_{ab}\}\,\tilde g^{mn} &=& i\,\partial_k\,\Big[\big\{B^k + \bar B^k + i\, (C^l \, \partial_l \, \bar C^k + \bar C^l \, \partial_l \, C^k)
\big\}\,\tilde g^{mn}\Big] \nonumber\\
&-& i\,\partial_k\,\Big[B^m + \bar B^m + i\, (C^l \, \partial_l \, \bar C^m + \bar C^l \, \partial_l \, C^m)\Big]\,\tilde g^{kn} \nonumber\\
&-& i\,\partial_k\,\Big[B^n + \bar B^n + i\, (C^l \, \partial_l \, \bar C^n + \bar C^l \, \partial_l \, C^n)\Big]\,\tilde g^{km}.
\end{eqnarray}
Thus, the off-shell nilpotent $[(s_{(a)b})^2 = 0]$ (anti-)BRST symmetry transformations [cf. Eqs. (7), (6)] are the {\it proper} set
of {\it quantum} symmetry transformations.

We end this section with the following remarks. First, the off-shell nilpotent $[s_{(a)b}^{2} = 0]$ (anti-)BRST symmetry transformations (7) and (6)
correspond to the {\it classical} 2D diffeomorphism symmetry transformations: $\xi^a \,\rightarrow \, g^{a}(\xi) = \xi^a - \varepsilon^a(\xi)$ 
where $g^{a}(\xi)$ is a physically well-defined function of $\xi^a$ on the 2D world-sheet such that it is finite at $\tau = 0$ and $\sigma = 0$ but vanishes 
off as $\tau \rightarrow  \pm \infty$ and $\sigma = \pi $. The infinitesimal version of these transformations 
are: $g^a(\xi) = \xi^a - \varepsilon^a(\xi)$ where $ \varepsilon^a(\xi)$ (with $a = 0, 1$) are the
2D infinitesimal diffeomorphism transformation parameters. Second, according to the basic tenets of BRST formalism, the parameters  $\varepsilon^a(\xi)$
have been replaced by the fermionic (anti-)ghost fields $(\bar C^a)C^a$ in the (anti-)BRST symmetry transformations (7) and (6). Third, it is crystal
clear, from Eq. (10), that the (anti-)BRST symmetry transformations $s_{(a)b}$ are absolutely anticommuting (i.e. $\{s_b, \, s_{ab}\} = 0$) in nature
only on the submanifold of the quantum Hilbert space of fields where the CF-type restrictions: $B^a + \bar B^a + i\, (C^m \, \partial_m \,
 \bar C^a + \bar C^m \, \partial_m \, C^a) = 0$ are satisfied. Finally, we note that the target space coordinates $X^\mu ({\xi})$ and
$[\mbox{det} \,\tilde g (\xi)]$ transform as {\it pure} Lorentz scalars [i.e. $ X^{\mu '} (\xi^{'} ) = X^{\mu}(\xi)$, $\,\mbox{det} 
\,\tilde g' (\xi ') =  \mbox{det} \,\tilde g (\xi) $] under the {\it infinitesimal} and continuous diffeomorphism symmetry transformations:
$\xi^a \,\rightarrow \, g^{a}(\xi) = \xi^a - \varepsilon^a(\xi)$.

\section {CF-Type Restrictions: MBTSA}

According to the basic tenets of MBTSA to BRST formalism, first of all, we generalize the 2D infinitesimal diffeomorphism transformations:
$\xi^{a} \longrightarrow \xi^{'a} = g^a(\xi) = \xi^a - \varepsilon^a(\xi)$ to its {\it counterpart} onto the $(2, 2)$-dimensional supermanifold as 
(see, e.g. [18, 13] for details)
\begin{eqnarray}
g^{a}(\xi) \quad \longrightarrow \quad \tilde g^{a}(\xi, \theta, \bar\theta) = \xi^a - \theta\,{\bar C}^a(\xi) - \bar\theta\,C^a(\xi) 
+ \theta\,\bar\theta\,f^a(\xi),
\end{eqnarray}
where the $(2, 2)$-dimensional supermanifold is parameterized by the superspace coordinates $Z^M = (\xi^a, \theta, \bar\theta)$. Here
$\xi^a = (\xi^0, \xi^1) \equiv (\tau, \sigma)$ are the bosonic world-sheet coordinates and a pair of Grassmannian variables $(\theta, \bar\theta)$
satisfy: $\theta^2 = {\bar\theta}^2 = 0$, $\theta\,\bar\theta + \bar\theta\,\theta = 0$. In Eq. (11), the {\it fermionic} 
(anti-)ghost $({\bar C}^a)C^a$ fields are the {\it ones} that are present in the 
(anti-)BRST transformations (7) and (6). In view of the mappings $(s_b \leftrightarrow \partial_{\bar\theta}\mid _{\theta = 0}, \,
s_{ab} \leftrightarrow \partial_{\theta}\mid _{\bar\theta = 0})$ established by Bonora and Tonin [4,5], the coefficients of $\theta$
and $\bar\theta$ in (11) have been taken to be the (anti-)ghost fields because, according to the standard BRST prescription, the
{\it classical} infinitesimal diffeomorphism symmetry transformations: $\delta\,\xi^a = -\, \varepsilon^a(\xi)$ have been promoted to the
{\it quantum} level by the (anti-)BRST symmetry transformations: $s_{ab}\,\xi^a = -\, {\bar C}^a, \, s_b\,\xi^a = -\, C^a$. The coefficients of
$\theta\,\bar\theta$ in (11) [i.e. $f^a(\xi)$] have to be determined from {\it other} consistency conditions of the BRST
formalism which we elaborate below.

To derive the CF-type restrictions and the (anti-)BRST symmetry transformations $s_{ab}\,X^\mu = {\bar C}^a\,\partial_a\,X^\mu, \, 
s_b\,X^\mu = C^a\,\partial_a\,X^\mu$, we generalize the target space {\it ordinary} coordinate fields $X^\mu(\xi)$ onto the $(2, 2)$-dimensional supermanifold
as   
\begin{eqnarray}
X^\mu(\xi) \quad \longrightarrow \quad {\tilde{X}}^\mu[\tilde g(\xi, \theta, \bar\theta), \theta, \bar\theta] &=& 
{\cal X}^\mu[\tilde g(\xi, \theta, \bar\theta)] + \theta\,{\bar R}^\mu[\tilde g(\xi, \theta, \bar\theta)] \nonumber\\
&+& \bar\theta\,
R^\mu[\tilde g(\xi, \theta, \bar\theta)] + \theta\,\bar\theta\,S^\mu[\tilde g(\xi, \theta, \bar\theta)], 
\end{eqnarray}
where ${\tilde X}^\mu\,[\tilde g(\xi, \theta, \bar\theta), \theta, \bar\theta]$ are the superfields whose arguments 
incorporate the super diffeomorphism transformations (11) and, on the r.h.s, we have the secondary superfields which have the following
super expansions [as their arguments are transformations (11)], namely;
\begin{eqnarray}
\theta \, \bar \theta \,S^\mu \,\big[\xi^a - \theta \, \bar C^a - \bar\theta \,C^a + \theta \,\bar\theta\, f^a  \big] &\equiv& 
\theta \, \bar \theta \,S^\mu (\xi^a) \equiv \theta \, \bar \theta \,S^\mu (\xi),\nonumber\\
\bar \theta \,R^\mu \,\big[\xi^a - \theta \, \bar C^a - \bar\theta \,C^a + \theta \,\bar\theta\, f^a  \big] &\equiv& 
\bar \theta \,R^\mu (\xi) + \theta \,  \bar \theta \, \bar C^a\,\partial_a\,R^{\mu}(\xi),\nonumber\\
\theta \,\bar R^\mu \,\big[\xi^a - \theta \, \bar C^a - \bar\theta \,C^a + \theta \,\bar\theta\, f^a  \big] &\equiv& 
\theta \,\bar R^\mu (\xi) - \theta \,  \bar \theta \,  C^a\,\partial_a\,\bar R^{\mu}(\xi),\nonumber\\
{\cal X}^\mu \,\big[\xi^a - \theta \, \bar C^a - \bar\theta \,C^a + \theta \,\bar\theta\, f^a  \big] &\equiv& 
 X^\mu (\xi) - \theta \, \bar C^a \, \partial_a\,X^\mu - \bar \theta \, C^a\,\partial_a\,X^\mu \nonumber\\
&+& \theta \,\bar\theta\,[f^a\,\partial_a\,X^\mu - \bar C^a \, C^m \partial_a\,\partial_m\,X^\mu], 
\end{eqnarray}
where ${\cal X}^\mu \,\big(\xi^a - \theta \, \bar C^a - \bar\theta \,C^a + \theta \,\bar\theta\, f^a  \big)|_{\theta\, = \,\bar \theta \,=\, 0}  =  X^\mu (\xi)$
and the Taylor expansions have been taken around $\theta\, = \,\bar \theta \,=\, 0$. Collecting the coefficients of $\theta$, $\bar \theta$
and $\theta \,\bar \theta$, from the r.h.s. of the {\it above} equation, we obtain the following:
\begin{eqnarray}
&&\tilde {X^\mu}\, [\tilde{g} ( \xi,\,\theta,\, \bar \theta ),\,\theta,\,\bar \theta ] = X^\mu ({\xi}) + \theta \,
[\bar R^\mu - \bar C^a\,\partial_a\,X^\mu ] + \bar \theta \,[ R^\mu -  C^a\,\partial_a\,X^\mu ]\nonumber\\ 
&&+ \theta \,\bar\theta \,[f^a\,\partial_a\,X^\mu - \bar C^a\,C^m\,\partial_a\,\partial_m\,X^\mu -
C^a\,\partial_a\,\bar R^\mu + \bar C^a \,\partial_a\,R^\mu + S^\mu].
\end{eqnarray}
We note that the target  space coordinate fields $X^\mu(\xi)$ are the {\it pure scalars} with respect to the 2D world-sheet on which we have taken
the diffeomorphism symmetry transformations $\xi^a \rightarrow \xi^{a}{'} = g^a (\xi)$. Thus, physically, it is evident that, ultimately, 
the restrictions on the (2, 2)-dimensional
superfield $\tilde {X^\mu}\, [\tilde{g} ( \xi,\,\theta,\, \bar \theta ),\,\theta,\,\bar \theta ]$ is the following:
\begin{eqnarray}
X^\mu(\xi) \rightarrow  \tilde {X^\mu}\, [\tilde{g} ( \xi,\,\theta,\, \bar \theta ),\,\theta,\,\bar \theta ] = X^\mu(\xi).
\end{eqnarray}
This is what has been called as the horizontality condition (HC) in [13, 18]. This HC [cf. Eq. (15)] amounts to setting the coefficients of 
$\theta$, $\bar \theta$ and $\theta \,\bar \theta$ in the expression (14) equal to zero. In other words, we have the following:
\begin{eqnarray}
&&R^\mu = C^a\,\partial_a\,X^\mu, \qquad  \bar R^\mu = \bar C^a\,\partial_a\,X^\mu, \nonumber\\
&&S^\mu = C^a\,\partial_a\,\bar R^\mu - \bar C^a\,\partial_a\,R^\mu + \bar C^a\,C^m\,\partial_a\,\partial_m\,X^\mu - f^a\,\partial_a\,X^\mu.
\end{eqnarray}
The last entry can be explicitly written by  plugging in the values of $R^\mu$ and $\bar R^\mu$ as 
\begin{eqnarray}
S^\mu = C^a\,\partial_a\,[\bar C_m\,\partial_m\,X^\mu] - \bar C^a\,\partial_a\,[C^m\,\partial_m\,X^\mu] + \bar C^a\,C^m\,\partial_a\,\partial_m\,X^\mu 
- f^a\,\partial_a\,X^\mu.
\end{eqnarray}
Now it is straightforward to check that we have the following:
\begin{eqnarray}
S^\mu = \big[C^a\,\partial_a\,\bar C^m - \bar C^a\,\partial_a\,C^m - f^m \big]\,(\partial_m\,X^\mu) - \bar C^m\,C^a\,\partial_m\,\partial_a\,X^\mu.
\end{eqnarray}
As pointed out earlier, the coefficients of $\theta\,\bar\theta$ [i.e. $f^a(\xi)$] in Eq. (11) and their presence in (18) can be computed 
by the requirements of the consistency conditions of BRST formalism.

One of the sacrosanct properties of a pure {\it scalar} field/superfield is the observation that {\it it} should {\it not}
transform {\it under} any kind of internal, spacetime, supersymmetric, etc., transformations. As a consequence, the secondary superfields of the r.h.s.
of (12) are 
\begin{eqnarray}
&&{\cal X}^\mu [\tilde g \, (\xi, \, \theta, \, \bar \theta )] = X^\mu (\xi), \qquad  \bar R^\mu 
[\tilde g \, (\xi, \, \theta, \, \bar \theta )] = \bar R^\mu (\xi),\nonumber\\
&&R^\mu [\tilde g \, (\xi, \, \theta, \, \bar \theta )] = R^\mu (\xi), \qquad  S^\mu [\tilde g \, (\xi, \, \theta, \, \bar \theta )] = S^\mu (\xi).
\end{eqnarray}
Similarly, the l.h.s. is: $\tilde X^\mu[\tilde g(\xi, \theta, \bar\theta), \theta, \bar\theta] = \tilde X^\mu(\xi, \theta, \bar\theta)$.
Substitutions of these equalities into (12) yield the following expressions in terms of $s_{(a)b}$, namely;
\begin{eqnarray}
{\tilde X}^\mu(\xi, \theta, \bar\theta) &=& X^\mu(\xi) + \theta\,{\bar R}^\mu(\xi) + \bar\theta\,R^\mu(\xi)
+ \theta\,\bar\theta \, S^\mu(\xi) \nonumber\\
&\equiv & X^\mu(\xi) + \theta\,(s_{ab}\,X^\mu) + \bar\theta\,(s_b\,X^\mu)
+ \theta\,\bar\theta \, (s_b\,s_{ab}\,X^\mu),
\end{eqnarray}
in a view of the Bonora-Tonin (BT) mappings: $s_b \leftrightarrow \partial_{\bar\theta}\mid _{\theta = 0}$, and 
$s_{ab} \leftrightarrow \partial_{\theta}\mid _{\bar\theta = 0}$ which was established in the realm of D-dimensional non-Abelian 1-form gauge theory [4,5]. 
In fact, a close look at (20) demonstrates that {\it this} expansion is exactly like the BT-superfield approach to BRST formalism in the 
context of gauge theories. Thus, it is clear from (16) and (18), that we have 
obtained the following [in terms of the (anti-)BRST symmetry transformations $(s_{(a)b})$ of (7) and (6)], namely;
\begin{eqnarray}
&&R^\mu = C^a\,\partial_a\,X^\mu = s_b\,X^\mu, \qquad  \bar R^\mu = \bar C^a\,\partial_a\,X^\mu = s_{ab}\,X^\mu, \nonumber \\
&&S^\mu = [C^a\,\partial_a\,\bar C^m - \bar C^a\,\partial_a\,C^m - f^m]\,(\partial_m\,X^\mu) - \bar C^a\,C^m\,\partial_a\,\partial_m\,X^\mu
\equiv  s_b\,s_{ab}\,X^\mu.
\end{eqnarray}
The absolute anticommutativity requirement (i.e. $\{s_b, s_{ab}\}\,X^\mu = 0$) implies  that the following 
equality is true, namely;
\begin{eqnarray}
s_b\,\bar R^\mu = -\,s_{ab}\,R^\mu \quad \Longleftrightarrow   \quad s_b\,s_{ab}\,X^\mu = -\,s_{ab}\,s_{b}\,X^\mu.
\end{eqnarray}
The explicit computations of $s_b\,\bar R^\mu$ and ($-\,s_{ab}\,R^\mu$) yield
\begin{eqnarray}
s_b\,\bar R^\mu &=& i\,B^m\,\partial_m\,X^\mu - \bar C^a\,C^m\,\partial_a\,\partial_m\,X^\mu - \bar C^a\,(\partial_a\,C^m)\,
(\partial_m\,X^\mu), \nonumber \\
-\,s_{ab}\,R^\mu &=& -\,i\,\bar B^m\,\partial_m\,X^\mu - \bar C^a\,C^m\,\partial_a\,\partial_m\,X^\mu + C^a\,(\partial_a\,\bar C^m)\,
(\partial_m\,X^\mu),
\end{eqnarray}
where we have used $s_b\,\bar C^a = i\,B^a$ and $s_{ab}\,C^a = i\,\bar B^a$. In addition, 
we have taken $s_b\,C^a = C^m\,\partial_m\,C^a$ and $s_{ab}\,\bar C^a = \bar C^m\,\partial_m\,
\bar C^a$ which are derived from the nilpotency requirements: $s_b^2\,X^\mu = 0$ and $s_{ab}^2\,X^\mu = 0$. 
The above equality (22) implies [from (23)] that we have
\begin{eqnarray}
B^m + \bar B^m + i\,(C^a\,\partial_a\,\bar C^m + \bar C^a\,\partial_a\,C^m) = 0,
\end{eqnarray}
which is nothing but the CF-type restrictions that have been obtained [cf. Eq. (10)] from the requirement of the absolute anticommutativity
property (i.e. $\{s_b, s_{ab}\} = 0$) of the (anti-)BRST symmetry transformations (7) and (6).

At this crucial juncture, we are in the position to determine the explicit expression for $f^a(\xi)$ that is present in  Eqs. (11) and (18)
by demanding the equality of {\it each} of the equations present in (23) with the expression for $S^\mu$ in (21). In other words, 
we find that:
\begin{eqnarray}
&&S^\mu = s_b\,\bar R^\mu \equiv -\,s_{ab}\,R^\mu \Longrightarrow \nonumber \\
&&\big[ C^a\,\partial_a\,\bar C^m - \bar C^a\,\partial_a\,C^m - f^m(\xi) \big]\,\partial_m\,X^\mu - \bar C^a\,C^m\,\partial_a\,\partial_m\,X^\mu
 \nonumber\\
&&= \big(i\,B^m - \bar C^a\,\partial_a\,C^m\big)\,\big(\partial_m\,X^\mu \big) - \bar C^a\,C^m\,\partial_a\,\partial_m\,X^\mu \nonumber\\
&&\equiv \big(-\,i\,\bar B^m + C^a\,\partial_a\,\bar C^m\big)\,\big(\partial_m\,X^\mu\big) - \bar C^a\,C^m\,\partial_a\,\partial_m\,X^\mu.
\end{eqnarray} 
A close look at (25) implies that there are {\it two} ways to equate the l.h.s [containing $f^m(\xi)$] with the r.h.s. of the 
above equation, namely;
\begin{eqnarray}
f^m (\xi) = -\,i\,B^m + \bar C^a\,\partial_a\,C^m  \equiv   i\,\bar B^m - C^a\,\partial_a\,\bar C^m,
\end{eqnarray}
which lead to the derivation of the CF-type restrictions (24). Thus, we conclude that the CF-type restrictions
are hidden in the determination of $f^a(\xi)$ of equation (11) by exploiting the absolute anticommutativity
property (i.e. $\{s_b, s_{ab}\}\,X^\mu = 0$) within the ambit of MBTSA to BRST formalism. Ultimately, we observe that
the above {\it logic} can be repeated  in the case of a pure {\it scalar}  $(\mbox{det}\, {\tilde g})$
to derive the CF-type restrictions (24)  and the (anti-)BRST transformations:
$s_{ab}\,(\mbox{det}\,\tilde g) = \bar C^a\,\partial_a\,(\mbox{det}\, \tilde g)$ and
$s_b\,(\mbox{det}\,\tilde g) = C^a\,\partial_a\,(\mbox{det}\, \tilde g)$, too.

We wrap-up this section with the following remarks. First of all, we have taken the standard (anti-)BRST symmetry transformations: 
$s_{ab}\,C^a = i\,\bar B^a, \, s_b\,\bar C^a = i\,B^a, \, s_{ab}\,\bar B^a = 0, \, s_b\,B^a = 0$ which imply the following [in the
terminology of the (anti-)chiral superfield approach (ACSA) to BRST formalism (see, e. g. [15])], namely;
\begin{eqnarray} 
C^m (\xi) \quad &\rightarrow& \quad F^{m\, (c)}_{(ab)} (\xi,\, \theta) = C^m(\xi) + \theta\,(i\,\bar B^m) \equiv C^m(\xi) + \theta\,(s_{ab}\,C^m), \nonumber\\
\bar C^m (\xi) \quad &\rightarrow& \quad \bar F^{m\, (ac)}_{(b)} (\xi,\, \bar\theta) = \bar C^m(\xi) + \bar\theta\,(i\,B^m) 
\equiv \bar C^m(\xi) + \bar\theta\, (s_{b}\,\bar C^m), \nonumber\\
B^m (\xi) \quad &\rightarrow& \quad \tilde B^{m\, (ac)}_{(b)} (\xi,\, \bar\theta) = B^m(\xi) + \bar\theta\,(0) 
\equiv B^m(\xi) + \bar\theta\, (s_{b}\,B^m), \nonumber\\
\bar B^m (\xi) \quad &\rightarrow& \quad \tilde{\bar B}^{m\, (c)}_{(ab)} (\xi,\, \theta) = \bar B^m(\xi) + \theta\,(0) \equiv \bar B^m(\xi) 
+ \theta\,(s_{ab}\,\bar B^m),
\end{eqnarray}
where the superscripts $(c)$ and $(ac)$ on the superfields [cf. the  l.h.s. of (27)] denote the 
{\it chiral} and {\it anti-chiral} versions of the {\it full} super expansions and the subscripts $(b)$ and $(ab)$ denote the fact
that the coefficients of $(\bar\theta)\,\theta$ in the above expansions lead to the determination of BRST and anti-BRST 
symmetry transformations. 
In other words, we are {\it sure} about the nilpotent (anti-)BRST symmetry transformations: $s_{ab}\,C^a = i\,\bar B^a, \, s_{ab}\,\bar B^a = 0, \, 
s_b\,\bar C^a = i\,B^a, \, s_b\,B^a = 0$ in terms of the {\it (anti-)chiral} superfield expansions in Eq. (27). Second, it is the off-shell
nilpotency requirements $s_{(a)b}^2\,X^\mu = 0$ which lead to $s_b\,C^a = C^m\,\partial_m\,C^a$ and $s_{ab}\,\bar C^a 
= \bar C^m\,\partial_m\,\bar C^a$. However, we have to obtain these transformations within the realm of superfield approach. Furthermore, it
is the requirement of the absolute anticommutativity properties: $\{s_b, s_{ab}\}\,C^a = 0, \, \{s_b, s_{ab}\}\,\bar C^a = 0$ which yield
$s_b\,\bar B^a = C^m\,\partial_m\,\bar B^a - \bar B^m\,\partial_m\,C^a$ and $s_{ab}\,B^a = \bar C^m\,\partial_m\,B^a - B^m\,\partial_m\,\bar C^a$.
We have to obtain, however, these symmetry transformations too, by using the techniques of the superfield approach to BRST formalism 
which we accomplish in our next section. Third,
we note that the  HC condition (15) has led to the following {\it full} super expansion of the target space coordinate superfield, namely;   
\begin{eqnarray}
\tilde X^{\mu\,(h)} (\xi,\, \theta,\,\bar\theta) &=& X^{\mu} (\xi) + \theta \,(\bar C^a\,\partial_a\,X^\mu) + \bar\theta\,(C^a\,\partial_a\,X^\mu)\nonumber\\
&+& \theta\,\bar\theta\,\big[(i\,B^a - \bar C^m\,\partial_m\,C^a)\,\partial_a\,X^\mu - \bar C^m\,C^a\,\partial_m\,\partial_a\,X^\mu\big]\nonumber\\
&\equiv & X^\mu (\xi) + \theta\,(s_{ab}\,X^\mu) + \bar\theta\,(s_b\,X^\mu) + \theta \,\bar\theta\,(s_b\,s_{ab}\,X^\mu),
\end{eqnarray}
where the superscript $(h)$ denotes the target space coordinate superfield that has been obtained after the application of HC which,
ultimately, leads to (20). Here the coefficients of $\theta$ and $\bar\theta$ are the (anti-)BRST symmetry transformations $[s_{(a)b}]$ 
that are listed in Eqs. (7) and (6). Finally, we comment that an expansion like (28) can be {\it also} written for the derivation of
the (anti-)BRST symmetry transformations for the {\it scalar} $(\mbox{det}\,\tilde g)$.

\section{(Anti-)BRST Symmetries of Other Fields: ACSA}

In this section, we exploit the theoretical strength of ACSA to BRST formalism (see, e. g. [15] and reference therein) 
to derive {\it all} the (anti-)BRST symmetry transformations
(7) and (6) {\it except} such transformations for the target space coordinates $X^\mu$ and   $(\mbox{det}\, {\tilde g})$ which have 
already been derived in the previous section by using MBTSA to BRST formalism [13, 18]. We are inspired to use, in our present section, ACSA  to BRST 
formalism because of our observations in Eq. (27). First of all, we focus on the derivation of the BRST symmetry transformations (7) which have {\it not}
been derived in the {\it previous} section. Thus, we wish to obtain: $s_b\,C^a = C^m\,\partial_m\,C^a, \, s_b\,\bar B^a = C^m\,\partial_m\,\bar B^a - 
\bar B^m\,\partial_m\,C^a, \, s_b\,{\tilde g}^{mn} = \partial_a\,(C^a\,\tilde g^{mn}) - (\partial_a\,C^m)\,\tilde g^{an} 
- (\partial_a\,C^n)\,{\tilde g}^{ma}, \, s_b\,E = (\partial_a\,C^a)\,E + C^a\,(\partial_a\,E)$. In this context, first of all, we
generalize the {\it ordinary} 2D fields $C^a(\xi), \, \bar B^a(\xi), \, E(\xi)$ and ${\tilde g}^{mn}(\xi)$ onto a $(2, 1)$-dimensional {\it anti-chiral} super
sub-manifold of the {\it general} $(2, 2)$-dimensional supermanifold as
\begin{eqnarray}
C^m(\xi) \quad &\rightarrow& \quad F^{m\,(ac)}(\xi, \bar\theta) = C^m(\xi) + \bar\theta\,b_1^m(\xi), \nonumber\\
\bar B^m(\xi) \quad &\rightarrow& \quad \bar{\cal B}^{m\,(ac)}(\xi, \bar\theta) = \bar B^m(\xi) + \bar\theta\,f_1^m(\xi), \nonumber\\
E(\xi) \quad &\rightarrow& \quad {\cal E}^{(ac)}(\xi, \bar\theta) = E(\xi) + \bar\theta\,f_2(\xi), \nonumber\\
\tilde g^{mn}(\xi) \quad &\rightarrow& \quad \tilde G^{mn\,(ac)}(\xi, \bar\theta) = \tilde g^{mn}(\xi) + \bar\theta\,\tilde R^{mn}(\xi), 
\end{eqnarray} 
where the 2D fields $(f_1^m, \, f_2, \, \tilde R^{ab})$ are {\it fermionic} secondary fields and $b_1^m(\xi)$ is a {\it bosonic} secondary field
due to the fermionic $(\bar\theta^2 = 0)$ nature of the Grassmannian variable $\bar\theta$. The above $(2, 1)$-dimensional {\it anti-chiral} super sub-manifold
is parameterized by $(\xi^a, \bar\theta)$ where $\xi^a \equiv (\tau, \sigma)$ are the {\it bosonic} coordinates and $\bar\theta$ is the 
fermionic $(\bar\theta^2 = 0)$ Grassmannian variable. The superscript $(ac)$ on the superfields denotes the {\it anti-chiral} super expansions of the
above {\it anti-chiral} superfields along $\bar\theta$-direction of the above super submanifold.

The basic tenets of ACSA to BRST formalism require that the BRST-invariant (i.e. {\it quantum} gauge invariant) quantities should be 
independent of the Grassmannian variables as the {\it latter} are {\it only} the mathematical artifacts that are useful in the context 
of theoretical techniques of SUSY theories. 
In this connection, we note that the following BRST (i.e. {\it quantum} gauge) invariant quantities are useful and important for us, namely;
\begin{eqnarray}
&&s_b\,\big[C^a\,\partial_a\,X^\mu\big] = 0, \qquad s_b\big[C^a\,\partial_a\,\bar B^m - \bar B^a\,\partial_a\,C^m\big] = 0, \nonumber\\
&&s_b\,\big[C^a\,\partial_a\,E + (\partial_a\,C^a)\,E\big] = 0, \nonumber\\
&&s_b\,\big[C^a\,\partial_a\,\tilde g^{mn} + (\partial_a\,C^a)\,\tilde g^{mn} - (\partial_a\,C^m)\,\tilde g^{an} - (\partial_a\,C^n)\,\tilde g^{ma}\big] = 0.
\end{eqnarray}
The above invariant quantities are obtained by a close observation of the transformations (7) where an off-shell nilpotency property 
$(s_b^2 = 0)$ exists for the BRST-symmetry transformations. We focus on $s_b\,[C^a\,\partial_a\,X^\mu] = 0$ which implies the following restriction
\begin{eqnarray}
F^{m\,(ac)}(\xi, \bar\theta)\,\partial_m\,X^{\mu\,(h, ac)}(\xi, \bar\theta) = C^m(\xi)\,\partial_m\,X^\mu(\xi),
\end{eqnarray} 
where $X^{\mu\,(h, ac)}(\xi, \bar\theta)$ is the {\it anti-chiral} limit of the {\it full} super expansion containing the nilpotent (anti-)BRST symmetries
as the coefficients of $\theta$ and $\bar\theta$. In other words, we have:
\begin{eqnarray}
X^{\mu\,(h, ac)}(\xi, \bar\theta) = X^\mu(\xi) + \bar\theta\,(C^a\,\partial_a\,X^\mu).
\end{eqnarray}
Plugging in the appropriate super expansions for $F^a(\xi, \bar\theta)$ from (29) as well as the super expansion for $X^{\mu\,(h, ac)}(\xi, \bar\theta)$ from
(32), we obtain the explicit expression for 
the secondary fields as: $b_1^m(\xi) = C^a\,\partial_a\,C^m$. As a consequence, we have the following {\it final} expansion
\begin{eqnarray}
F^{m\,(ac)}_{(b)}(\xi, \bar\theta) = C^m(\xi) + \bar\theta\,(C^a\,\partial_a\,C^m) \equiv C^m(\xi) + \bar\theta\,(s_b\,C^m),
\end{eqnarray} 
where the subscript $(b)$ on the superfield (on the l.h.s.) denotes that the above {\it anti-chiral} superfield has been obtained 
after the application of the BRST 
invariant restrictions (31) and the coefficient of $\bar\theta$ is nothing but the BRST symmetry transformation for the field $C^m(\xi)$
which also encodes the relationships: $\partial_{\bar\theta}\,F^{m\,(ac)}(\xi, \bar\theta) = s_b\,C^m(\xi)$ and $\partial_{\bar\theta}^2 = 0
\Leftrightarrow s_b^2 = 0$. The {\it latter} establishes the connection between the nilpotency properties of $\partial_{\bar\theta}$ and $s_b$.

At this juncture, we now concentrate on the derivation of $f_2(\xi)$ in the expansion of ${\cal E}^{ac}(\xi, \bar\theta)$ in Eq. (29).
For this purpose, we note that $s_b\,[C^m\,\partial_m\,E + (\partial_m\,C^m)\,E] = 0$. Following the basic principle of ACSA, the expressions 
in the square bracket have to be generalized onto the $(2, 1)$-dimensional {\it anti-chiral} super sub-manifold with the following 
BRST (i.e. {\it quantum} gauge) symmetry invariant restriction
\begin{eqnarray}
F_{(b)}^{m\,(ac)}(\xi, \bar\theta)\,\partial_m\,{\cal E}^{(ac)}(\xi, \bar\theta) + \big[\partial_m\,F_{(b)}^{m\,(ac)}(\xi, \bar\theta)\big]\,
{\cal E}^{(ac)}(\xi, \bar\theta) \nonumber\\ 
= C^m(\xi)\,\big[\partial_m\,E(\xi)\big] + \big[\partial_m\,C^m(\xi)\big]\,E(\xi),
\end{eqnarray}
where the expansions of $F_{(b)}^{m\,(ac)}(\xi, \bar\theta)$ and ${\cal E}^{(ac)}(\xi, \bar\theta)$ have been quoted in Eqs. (33) and (29), respectively.
Substitutions of these super expansions into the l.h.s. and comparison with the r.h.s. of the restriction (34), lead to the following condition 
\begin{eqnarray}
&&(\partial_m\,C^a)\,(\partial_a\,C^m)\,E + C^a\,(\partial_a\,\partial_m\,C^m)\,E + C^a\,(\partial_a\,C^m)\,(\partial_m\,E)\nonumber\\ 
&&- (\partial_m\,C^m)\,f_2 
- C^m\,(\partial_m\,f_2) = 0.
\end{eqnarray}
In other words, the restriction (34) implies that BRST invariant quantity {\it must} be independent of $\bar\theta$. A careful and close look
at the above equation leads to:
\begin{eqnarray}
\partial_m\,\big[C^a\,(\partial_a\,C^m)\,E - C^m\,f_2\big] = 0.
\end{eqnarray}
Substituting for $C^a\,(\partial_a\,C^m)\,E = \partial_a\,[C^a\,C^m\,E] -\, (\partial_a \,C^a)\, C^m \,E - \,C^a\,C^m\,(\partial_a \,E),$ we obtain the following
from the above equation:
\begin{eqnarray}
\partial_m \, [\partial_a\,\{C^a\,C^m\,E\} -\, (\partial_a \,C^a)\,C^m\, E - C^a\,C^m\,(\partial_a\,E) - C^m\,f_2] = 0.
\end{eqnarray}
It is clear that the {\it first} term in the square-bracket will be {\it zero} if we operate the derivative $(\partial_m)$ from outside. Thus, 
the final expression is as follows:
\begin{eqnarray}
\partial_m\,[C^m\,\{\partial_a\,(C^a\,E) - f_2\}] = 0.
\end{eqnarray} 
Integrating over $d^2\,\xi = d\,\sigma\,d\,\tau$ and taking the physicality condition that all the fields {\it must} vanish off as
$\tau \rightarrow \pm\,\infty$ {\it and} at $\sigma = 0$, $\sigma = \pi$, we obtain the precise value of $f_2(\xi)$ as
\begin{eqnarray}
f_2 = \partial_a\,(C^a\,E) \qquad [\mbox{for}\,\, C^m \ne 0].
\end{eqnarray}
Hence, we have the following {\it final} expansion for the superfield ${\cal E}^{(ac)}(\xi, \bar\theta)$
\begin{eqnarray}
{\cal E}_{(b)}^{(ac)}(\xi, \bar\theta) = E(\xi) + \bar\theta\,\big[\partial_n(C^n\,E)\big] \equiv E(\xi) + \bar\theta\,(s_b\,E),
\end{eqnarray}
which leads to the derivation of the BRST symmetry transformation $s_b\,E = \partial_a\,(C^a\,E)$ as the coefficient of $\bar\theta$ in the above equation
implying, once again, that $\partial_{\bar\theta}\,{\cal E}_{(b)}^{(ac)}(\xi, \bar\theta) = s_b\,E(\xi)$. This relationship establishes the connection
between $s_b$ and translational generator $\partial_{\bar\theta}$ along the $\bar\theta$-direction of the (2, 1)-dimensional {\it anti-chiral}
super sub-manifold and it also demonstrates that $s_b^2 = 0 \Leftrightarrow \partial_{\bar\theta}^2 = 0$ (which is the connection between the 
nilpotency properties). It goes without saying that the subscript $(b)$ on the l.h.s. denotes that the super expansion (40) has been
obtained after the application of the BRST invariant restriction (34).

We now focus on the BRST invariance: $s_b\,[C^n\,\partial_n\,\bar B^m - \bar B^n\, \partial_n\,C^m] = 0$. This 
observation can be generalized onto the (2, 1)-dimensional {\it anti-chiral} super
sub-manifold with the following restriction on the {\it anti-chiral} superfields, namely; 
\begin{eqnarray}
&&F_{(b)}^{m\,(ac)}(\xi, \bar\theta)\,\partial_m\,{\bar{\cal B}}^{n\,(ac)}(\xi, \bar\theta) 
- {\bar{\cal B}}^{m\,(ac)}(\xi, \bar\theta)\,\partial_m\,F_{(b)}^{n\,(ac)}(\xi, \bar\theta)\nonumber\\
&& = C^m(\xi)\,\partial_m\,\bar B^n (\xi) - \bar B^m(\xi)\,
\partial_m\,C^n(\xi).
\end{eqnarray}
The substitutions of expansions from (29) and (33) lead to the following equality:
\begin{eqnarray}
C^n\,\big[ \partial_n\,f^{m}_{1} + {\bar B}^a\,(\partial_a\,\partial_n\,C^m) - (\partial_n\,C^a)(\partial_a\,{\bar B}^m)\big]
+ \big[f^{a}_{1} + {\bar B}^n\,(\partial_n\,C^a)\big](\partial_a\,C^m) = 0.
\end{eqnarray}
In the above, the term: $ -\,(\partial_n\,C^a)(\partial_a\,{\bar B}^m)$ can be 
written as: $  -\, \partial_n\,[C^a\,\partial_a\,{\bar B}^m] 
+ C^a\,\partial_n \,\partial_a\,{\bar B}^m$. It is elementary to note that the {\it second term} will vanish-off when
we shall multiply by $ C^n$ from the left (i.e. $C^n\, C^a\, \partial_a \, \partial_n\, \bar B^m = 0$).
The substitution of the {\it leftover} term (i.e. $ -\, \partial_n\,[C^a\,\partial_a\,{\bar B}^m]  $) into (42) leads to: 
\begin{eqnarray}
C^n\,\partial_n\,\big[f^{m}_{1} + {\bar B}^a\,(\partial_a\,C^m) - C^a\,\partial_a\,{\bar B }^m\big] + \big[f^{a}_{1} 
+ {\bar B}^n\,(\partial_n\,C^a) - C^n\,\partial_n\,{\bar B}^a\big] (\partial_a\,C^m) = 0.
\end{eqnarray}
It is straightforward to note that $f^{m}_{1} = C^a\,\partial_a\,{\bar B }^m - {\bar B}^a\,(\partial_a\,C^m)$ satisfies 
the above equation very beautifully. Thus, we have, ultimately, the following expansion [cf. Eq. (29)]:  
\begin{eqnarray}
\bar{\cal B}^{m\,(ac)}_{(b)}(\xi, \bar\theta) = \bar B^m(\xi) + \bar\theta\,[C^a\,\partial_a\,{\bar B }^m - {\bar B}^a\,\partial_a\,C^m]
\equiv \bar B^m(\xi) + \bar \theta \,[s_b\,\bar B^m(\xi)].
\end{eqnarray}
Hence , we have derived the BRST transformations: $s_b\,\bar B^m = C^a\,\partial_a\,{\bar B }^m - {\bar B}^a\,\partial_a\,C^m$ as the coefficient of 
$\bar\theta $ in the above super expansion. It should be noted that the subscript $(b)$ on the superfield [cf. l.h.s. of Eq. (44)] denotes
that $\bar{\cal B}^{m\,(ac)}_{(b)}(\xi, \bar\theta)$ has been derived after the imposition of the BRST invariant restriction (41).

At this stage, we now wish to derive the BRST symmetry transformation $[s_b\,{\tilde g}^{mn} = \partial_k (C^k \, {\tilde g}^{mn}) \, - \, 
(\partial_k\,C^m) \,{\tilde g}^{kn} \, - \, (\partial_k \, C^n)\, {\tilde g}^{mk}]$ using the theoretical strength of ACSA to 
BRST formalism. Towards this goal in mind, we have the following restriction on the {\it anti-chiral} superfields which 
have their super expansions in (29) and (33), namely;
\begin{eqnarray}
&& F^{k(ac)}_{(b)}(\xi, \bar\theta)\,\partial_k\,\tilde G^{mn(ac)}(\xi, \bar\theta) + \big[\partial_k\,F^{k(ac)}_{(b)}(\xi, \bar\theta)\big]\,
\tilde G^{mn(ac)}(\xi, \bar\theta) \nonumber\\
&& - \big[\partial_k\,F^{m(ac)}_{(b)}(\xi, \bar\theta)\big]\,\tilde G^{kn(ac)}(\xi, \bar\theta) 
 - \big[\partial_k\,F^{n(ac)}_{(b)}(\xi, \bar\theta)\big]\,\tilde G^{km(ac)}(\xi, \bar\theta) \nonumber\\
&& = C^k(\xi)\,\big[\partial_k\,\tilde g^{mn}(\xi)\big] + \big[\partial_k\,C^k(\xi)\big]\,\tilde g^{mn}(\xi) \nonumber\\
&& - \big[\partial_k\,C^m(\xi)\big]\,\tilde g^{kn}(\xi) - \big[\partial_k\,C^n(\xi)\big]\,\tilde g^{mk}(\xi).
\end{eqnarray}
The above restriction has been obtained by a close look at the off-shell nilpotency property $(s_b^2\,\tilde g^{mn} = 0)$ of 
the BRST symmetry transformations (7). This 
restriction on the {\it anti-chiral} superfields leads to the following condition on the {\it basic} and {\it secondary} fields
\begin{eqnarray}
&& C^k\,(\partial_k\,\tilde R^{mn}) + (\partial_k\,C^k)\,\tilde R^{mn} -(\partial_k\,C^l)\,(\partial_l\,C^k)\,\tilde g^{mn} 
- C^l\,(\partial_k\,\partial_l\,C^k)\,\tilde g^{mn} \nonumber\\
&& - C^l\,(\partial_l\,C^k)\,(\partial_k\,\tilde g^{mn}) - (\partial_k\,C^m)\,\tilde R^{kn} + (\partial_k\,C^l)\,(\partial_l\,C^m)\,\tilde g^{kn} 
+ C^l\,(\partial_k\,\partial_l\,C^m)\,\tilde g^{kn} \nonumber\\
&& - (\partial_k\,C^n)\,\tilde R^{mk} + (\partial_k\,C^l)\,(\partial_l\,C^n)\,\tilde g^{mk} + C^l\,(\partial_k\,\partial_l\,C^n)\,\tilde g^{mk} = 0,   
\end{eqnarray}
where we have used the super expansions from (29) and (33). It is straightforward to note that the first {\it five} 
terms, in the above, lead to the following total derivative, namely;
\begin{eqnarray}
\partial_k\,\big[C^k\,\tilde R^{mn} - C^l\,(\partial_l\,C^k)\,\tilde g^{mn}\big] \equiv \partial_k\,\big[C^k\,\{\tilde R^{mn} 
- \partial_l\,(C^l\,\tilde g^{mn})\}\big],
\end{eqnarray}
where we have used: $-\,C^l\,(\partial_l\,C^k)\,\tilde g^{mn} = -\,\partial_l\,\big[C^l\,C^k\,\tilde g^{mn}\big] + (\partial_l\,C^l)\,\tilde g^{mn}
+ C^l\,C^k\,(\partial_l\,\tilde g^{mn})$ and $\partial_k\,\partial_l\,(C^l\,C^k\,\tilde g^{mn}) = 0$. Adding and substracting: $\partial_k\,
\big[C^k\,(\partial_l\,C^m)\,\tilde g^{ln} + C^k\,(\partial_l\,C^n)\,\tilde g^{ml}\big]$ we obtain the following equation from (46):
\begin{eqnarray}
&& \partial_k\,\big[C^k\,\{\tilde R^{mn} - \partial_l\,(C^l\,\tilde g^{mn}) + (\partial_l\,C^m)\,\tilde g^{ln} 
+ (\partial_l\,C^n)\,\tilde g^{ml}\}\big] \nonumber\\
&& - \partial_k\,\big[C^k\,(\partial_l\,C^m)\,\tilde g^{ln} + C^k\,(\partial_l\,C^n)\,\tilde g^{ml}\big] = 0.
\end{eqnarray}
Expanding the {\it total} derivative in the {\it second} entry of the above equation and rearranging these, we obtain the following
interesting equation, namely;
\begin{eqnarray}
&& \partial_k\,\big[C^k\,\{\tilde R^{mn} - \partial_l\,(C^l\,\tilde g^{mn}) + (\partial_l\,C^m)\,\tilde g^{ln} + (\partial_l\,C^n)\,
\tilde g^{ml}\}\big] \nonumber\\
&& - (\partial_k\,C^m)\,\big[\tilde R^{nk} - \partial_l\,(C^l\,\tilde g^{nk}) + (\partial_l\,C^k)\,\tilde g^{ln} \big] \nonumber\\
&& - (\partial_k\,C^n)\,\big[\tilde R^{mk} - \partial_l\,(C^l\,\tilde g^{mk}) + (\partial_l\,C^k)\,\tilde g^{lm} \big] = 0.  
\end{eqnarray}
Adding and subtracting $(\partial_k\,C^m)\,(\partial_l\,C^n)\,\tilde g^{lk} + (\partial_k\,C^n)\,(\partial_l\,C^m)\,\tilde g^{lk}$ in the above,
we finally obtain the following very nice looking equation:
\begin{eqnarray}
&& \partial_k\,\big[C^k\,\{\tilde R^{mn} - \partial_l\,(C^l\,\tilde g^{mn}) + (\partial_l\,C^m)\,\tilde g^{ln} + (\partial_l\,C^n)\,
\tilde g^{ml}\}\big] \nonumber\\
&& - (\partial_k\,C^m)\,\big[\tilde R^{nk} - \partial_l\,(C^l\,\tilde g^{nk}) + (\partial_l\,C^k)\,\tilde g^{ln} 
+ (\partial_l\,C^n)\,\tilde g^{lk} \big] \nonumber\\
&& - (\partial_k\,C^n)\,\big[\tilde R^{mk} - \partial_l\,(C^l\,\tilde g^{mk}) + (\partial_l\,C^k)\,\tilde g^{lm} + (\partial_l\,C^m)\,\tilde g^{lk}\big] = 0.  
\end{eqnarray}   
It should be noted that what we have {\it added} and subtracted in (49) is {\it basically} equal to {\it zero} on its own because we make the following
observation:
\begin{eqnarray}
\tilde g^{lk}\,\big[(\partial_k\,C^m)\,(\partial_l\,C^n) + (\partial_k\,C^n)\,(\partial_l\,C^m)\big] = 0.
\end{eqnarray}
In other words, the {\it last} entries in the {\it second} and {\it third} lines of Eq. (50) are zero {\it on}
their {\it own}. We note that the {\it symmetric} indices in $(\tilde g^{lk})$ and {\it anti-symmetric} indices $(l, k)$ in 
the square-bracket are sum-up to yield {\it zero}. It is straightforward now to point out that
\begin{eqnarray}
\tilde R^{mn} = \partial_k\,(C^k\,\tilde g^{mn}) - (\partial_k\,C^m)\,\tilde g^{kn} - (\partial_k\,C^n)\,\tilde g^{mk},  
\end{eqnarray}   
satisfies the above equation (50). As a consequence, we have the following
\begin{eqnarray}
\tilde G^{mn(ac)}_{(b)}(\xi, \bar\theta) &=& \tilde g^{mn}(\xi) + \bar\theta\,\big[\partial_k\,(C^k\,\tilde g^{mn}) - (\partial_k\,C^m)\,\tilde g^{kn} 
- (\partial_k\,C^n)\,\tilde g^{mk}\big] \nonumber\\
& \equiv & \tilde g^{mn}(\xi) + \bar\theta\,[s_b\,\tilde g^{mn}(\xi)],
\end{eqnarray}
where the coefficient of $\bar\theta$ is nothing but the BRST symmetry transformation for $\tilde g^{mn}(\xi)$ that has been 
quoted in (7). The subscript $(b)$ on the l.h.s. of the above equation denotes that the {\it anti-chiral} superfield 
$\tilde G^{mn}_{(b)}(\xi, \bar\theta)$ has been obtained after the application of the BRST invariant restriction on a
specific combination of superfields [cf. Eq. (45)].

We set out {\it now} to derive the anti-BRST symmetry transformations (6) by using ACSA to BRST formalism where, first of all, we generalize
the following {\it basic} and {\it auxiliary} fields of our theory onto a $(2, 1)$-dimensional {\it chiral} super submanifold
\begin{eqnarray}
B^m(\xi) \quad &\rightarrow& \quad {\cal B}^{m(c)}(\xi, \theta) = B^{m}(\xi) + \theta\,\bar f_1^{m}(\xi), \nonumber\\
E(\xi) \quad &\rightarrow& \quad {\cal E}^{(c)}(\xi, \theta) = E(\xi) + \theta\,\bar f_2(\xi), \nonumber\\
\bar C^m(\xi) \quad &\rightarrow& \quad {\bar F}^{m(c)}(\xi, \theta) = \bar C^{m}(\xi) + \theta\,\bar b_1^{m}(\xi), \nonumber\\
\tilde g^{mn}(\xi) \quad &\rightarrow& \quad \tilde G^{mn(c)}(\xi, \theta) = \tilde g^{mn}(\xi) + \theta\,\tilde{\bar R}^{mn}(\xi),
\end{eqnarray}
where $(\bar f_1^{m},\,\bar f_2^{m},\,\tilde{\bar R}^{mn})$ are the {\it fermionic} and $\bar b^{m}_{1}$ is the {\it bosonic}
secondary fields that are to be determined in terms of the {\it basic} and {\it auxiliary} fields of the (anti-)BRST
invariant Lagrangian densities ${\cal L}_{(a)b}$ [cf. Eqs. (1), (2)]. It is elementary to note that, in the limit $\theta = 0$,
we retrieve the {\it bosonic} and {\it auxiliary} fields of ${\cal L}_{(a)b}$. We point out that $s_{ab}\,\bar B^m(\xi) = 0$
implies that we have $\bar{\cal B}^m_{(ab)} (\xi,\, \theta) = \bar B^m (\xi)$ where $\bar{\cal B}^m_{(ab)} (\xi,\, \theta)$ 
is the superfield that has been obtained after the restriction on the {\it chiral} superfield $\bar{\cal B}^m (\xi,\, \theta) $
that is obtained in the generalization $\bar B^m(\xi)\rightarrow \bar{\cal B}^m (\xi,\, \theta)$ on the {\it chiral} 
super submanifold [which is parameterized by $(\xi^a,\, \theta)$ where $\xi^a$ characterize the 2D world-sheet and $\theta$
is the fermionic $(\theta^2 = 0)$ Grassmannian variable]. The subscript $(ab)$ denotes the {\it chiral} superfield which leads to
the derivation of $[s_{ab}\, \bar B (\xi) = 0]$ as the coefficient of $\theta$ in its expansion: 
$\bar{\cal B}^m_{(ab)} (\xi,\, \theta) = \bar B^m (\xi) + \theta\,(0)\,\equiv \,\bar B^m (\xi) + \theta \,(s_{ab}\,\bar B^m)$.
It should be further noted that we have {\it not} devoted time on the derivation of the (anti-)BRST symmetries that have
already been derived and mentioned in Sec. 3 where the theoretical strength of MBTSA has been exploited.

A close and careful observation of the anti-BRST symmetry transformations (6) demonstrates that we have the following very
{\it useful} and interesting combinations of fields 
\begin{eqnarray}
&&s_{ab}\,\big[\bar C^a\,\partial_a\,X^\mu\big] = 0, \qquad s_{ab}\big[\bar C^a\,\partial_a\,B^m - B^a\,\partial_a\,\bar C^m\big] = 0, \nonumber\\
&&s_{ab}\,\big[\bar C^a\,\partial_a\,E + (\partial_a\,\bar C^a)\,E\big] = 0, \nonumber\\
&&s_{ab}\,\big[\bar C^a\,\partial_a\,\tilde g^{mn} + (\partial_a\,\bar C^a)\,\tilde g^{mn} - (\partial_a\,\bar C^m)\,\tilde g^{an} 
- (\partial_a\,\bar C^n)\,\tilde g^{ma}\big] = 0,
\end{eqnarray}
as the anti-BRST invariant quantities. The fundamental requirement
of ACSA is that the generalizations of the quantities [present in the square bracket of (55)] onto a suitably chosen $(2, 1)$-dimensional {\it chiral}
super submanifold should be independent of the Grassmannian variable $\theta$. As a consequence, we have the following restrictions
\begin{eqnarray} 
&& \bar F^{a(c)}(\xi, \theta)\,\partial_a\,X^{\mu(h, c)}(\xi, \theta) = \bar C^a(\xi)\,\partial_a\,X^{\mu}(\xi), \nonumber\\
&&\bar F^{a\,(c)}(\xi, \theta)\,\partial_a\,{{\cal B}}^{m\,(c)}(\xi, \theta) 
- {{\cal B}}^{a\,(c)}(\xi, \theta)\,\partial_a\,\bar F^{m\,(c)}(\xi, \theta)
 = \bar C^a(\xi)\,\partial_a\,B^m (\xi) - B^a(\xi)\,\partial_a\,\bar C^m(\xi), \nonumber\\
&& \bar F^{a(c)}(\xi, \theta)\,\partial_a\,{\cal E}^{(c)}(\xi, \theta) + \big[\partial_a\,\bar F^{a(c)}(\xi, \theta)\big]\,
{\cal E}^{(c)}(\xi, \theta) 
= \bar C^a(\xi)\,\partial_a\,E(\xi) + \big[\partial_a\,\bar C^a(\xi)\big]\,E(\xi), \nonumber\\
&& \bar F^{a(c)}(\xi, \theta)\,\partial_a\,\tilde G^{mn(c)}(\xi, \theta) + \big[\partial_a\,\bar F^{a(c)}(\xi, \theta)\big]\,
\tilde G^{mn(c)}(\xi, \theta) \nonumber\\
&& - \big[\partial_a\,\bar F^{m(c)}(\xi, \theta)\big]\,\tilde G^{an(c)}(\xi, \theta) 
 - \big[\partial_a\,\bar F^{n(c)}(\xi, \theta)\big]\,\tilde G^{ma(c)}(\xi, \theta) \nonumber\\
&& = \bar C^a(\xi)\,\big[\partial_a\,\tilde g^{mn}(\xi)\big] + \big[\partial_a\,\bar C^a(\xi)\big]\,\tilde g^{mn}(\xi) 
 - \big[\partial_a\,\bar C^m(\xi)\big]\,\tilde g^{an}(\xi) - \big[\partial_a\,\bar C^n(\xi)\big]\,\tilde g^{ma}(\xi),
\end{eqnarray}
where we have taken the super expansions from (54) and $X^{\mu(h, c)}(\xi, \theta)$ is the {\it chiral} limit $(\bar\theta = 0)$
of the {\it full} expansion [cf. (28)]. In other words, we have the following
\begin{eqnarray}
X^{\mu(h, c)}(\xi, \theta) = X^{\mu}(\xi) + \theta\,[\bar C^a\,\partial_a\,X^{\mu}(\xi)],
\end{eqnarray}
where the superscript $(h, c)$ denotes the {\it chiral} version of the {\it full} expansion of $X^{\mu(h)}(\xi, \theta)$ that has
been obtained in the previous section [cf. Eq. (28)].

We would like to lay emphasis on the fact that {\it all} the secondary fields $(\bar f_1^m,\,f_2,\,\tilde{\bar R}^{mn})$ and $\bar b_1^m$
can be computed in {\it exactly} similar manner as we have done in the case of determination of the BRST symmetry transformations $(s_b)$
for the super expansions in Eq. (29). It turns out that, adopting {\it this} logic, we obtain the following:
\begin{eqnarray} 
&& \bar f^m_1 = \bar B^a\,\partial_a\,\bar C^m - (\partial_a\,\bar B^a)\,\bar C^m, \qquad f_2 = (\partial_a\,\bar C^a)\,E 
+ \bar C^a\,(\partial_a\,E), \nonumber\\
&& \bar b_1^m = \bar C^a\,\partial_a\,\bar C^m,  \qquad
 \tilde{\bar R}^{mn} = \partial_a\,(\bar C^a\,\tilde g^{mn}) - (\partial_a\,\bar C^m)\,\tilde g^{an} - (\partial_a\,\bar C^n)\,\tilde g^{ma}.
\end{eqnarray}
Substitutions of the above secondary fields into the {\it chiral} super expansions of Eq. (54), we obtain the following
{\it final} super expansions
\begin{eqnarray}
{\cal B}^{m(c)}_{(ab)}(\xi, \theta) &=& B^m(\xi) + \theta\,\big[\bar C^a\,\partial_a\,B^m - B^a\,\partial_a\,\bar C^m\big] \equiv B^m(\xi) 
+ \theta\,\big[s_{ab}\,B^m(\xi)\big], \nonumber\\
{\cal E}^{(c)}_{(ab)}(\xi, \theta) &=& E(\xi) + \theta\,\big[\partial_a\,(\bar C^a\,E)\big] \equiv E(\xi) 
+ \theta\,\big[s_{ab}\,E(\xi)\big], \nonumber\\
{\bar F}^{m(c)}_{(ab)}(\xi, \theta) &=& \bar C^m(\xi) + \theta\,\big[\bar C^a\,\partial_a\,\bar C^m\big] \equiv \bar C^m(\xi) 
+ \theta\,\big[s_{ab}\,\bar C^m(\xi)\big], \nonumber\\
{\tilde G}^{mn(c)}_{(ab)}(\xi, \theta) &=& \tilde g^{mn}(\xi) + \theta\,\big[\partial_a\,(\bar C^a\,\tilde g^{mn}) 
- (\partial_a\,\bar C^m)\,\tilde g^{an} - (\partial_a\,\bar C^n)\,\tilde g^{ma}\big] \nonumber\\ 
&\equiv& \tilde g^{mn}(\xi) + \theta\,\big[s_{ab}\,\tilde g^{mn}(\xi)\big], 
\end{eqnarray}
where the subscript $(ab)$ on the {\it chiral} superfields on the l.h.s. of the above equation (59) denotes that the {\it above} 
superfields have been obtained after the {\it quantum} gauge (i.e. anti-BRST) invariant restrictions on the {\it chiral} superfields
[cf. Eq. (56)] have been imposed. It can be readily checked that we have obtained the anti-BRST symmetry transformations:
$s_{ab}\,B^m = \bar C^a\,\partial_a\,B^m - B^a\,\partial_a\,\bar C^m, \, s_{ab}\,E = \partial_a\,[\bar C^a\,E], \, 
s_{ab}\,\bar C^m = \bar C^a\,\partial_a\,\bar C^m, \, s_{ab}\,\tilde g^{mn} = \partial_a\,(\bar C^a\,\tilde g^{mn}) 
- (\partial_a\,\bar C^m)\,\tilde g^{an} - (\partial_a\,\bar C^n)\,\tilde g^{ma}$ as the coefficients of the chiral super
expansions in (59). It is nice to note that $\partial_{\theta}\,\Omega_{(ab)}(\xi, \theta) = s_{ab}\,\omega(\xi)$ where
the {\it generic} chiral superfield $\Omega_{(ab)}(\xi, \theta)$ stands for the l.h.s. of (59) and $\omega = B^m, \, E,\, \bar C^m, \,
\tilde g^{mn}$ generic {\it ordinary} field.

We end this section with the following remarks. First, we have derived the (anti-)BRST symmetry transformations for the fields by exploiting the 
theoretical tricks of ACSA to BRST formalism. These fields are the {\it ones} for which the MBTSA has {\it not} been able to derive the 
(anti-)BRST symmetry transformations. Second, a careful and close observation of the theoretical contents of Secs. 3 and 4 demonstrate
that we have derived {\it all} the nilpotent (anti-)BRST symmetry transformations for our theory by exploiting the theoretical strength of MBTSA and ACSA.
Finally, the (anti-)BRST symmetry transformations for the component fields $A_0,\,A_1$ and $A_2$ of $\tilde g^{mn}$ [cf. Eq. (4)] 
can be obtained from the {\it exact} expressions for $ s_b \,\tilde g^{mn} (\xi)$ and $ s_{ab} \,\tilde g^{mn} (\xi)$ that
have been quoted in (7) and (6). To be more transparent, we find the
following {\it anti-chiral} super expansions:
\begin{eqnarray*}  
A_0(\xi) \quad \rightarrow \quad {\cal A}^{(ac)}_{0(b)}(\xi, \bar\theta) &=& A_0(\xi) + \bar\theta\,\big[ C^m\,\partial_m\, A_0  - 
(\partial_0 \, C^1 - \partial_1 \,C^0) \, A_2 \nonumber\\
&-& (\partial_0 \, C^1 + \partial_1 \, C^0)\,  A_1\big]  \equiv A_0(\xi) + \bar\theta\,\big[s_b\,A_0(\xi)\big], \nonumber\\
\end{eqnarray*}
\begin{eqnarray} 
A_1(\xi) \quad \rightarrow \quad {\cal A}^{(ac)}_{1(b)}(\xi, \bar\theta) &=& A_1(\xi) + \bar\theta\,\big[ C^m\,\partial_m \, A_1  - 
(\partial_1 \,C^0 + \partial_0 \, C^1)\, A_0 \nonumber\\
&-& (\partial_0 \, C^0 - \partial_1 \, C^1) \, A_2\big]  \equiv A_1(\xi) + \bar\theta\,\big[s_b\,A_1(\xi)\big], \nonumber\\
A_2(\xi) \quad \rightarrow \quad {\cal A}^{(ac)}_{2(b)}(\xi, \bar\theta) &=& A_2(\xi) + \bar\theta\,\big[C^m\,\partial_m \, A_2 - 
(\partial_1 \, C^0 - \partial_0 \, C^1) A_0 \nonumber\\
&-& (\partial_0 \, C^0 - \partial_1 \, C^1) A_1\big]  \equiv A_2(\xi) + \bar\theta\,\big[s_b\,A_2(\xi)\big], 
\end{eqnarray}
where the coefficients of $\bar\theta$ are nothing but the BRST symmetry transformations [cf. Eq. (9)] on $A_0(\xi), \, A_1(\xi)$
and $A_2(\xi)$. In exactly similar fashion, we can obtain the anti-BRST symmetry transformations on $A_0, \, A_1$ and $A_2$ from
the following {\it chiral} super expansions:
\begin{eqnarray}  
A_0(\xi) \quad \rightarrow \quad {\cal A}^{(c)}_{0(ab)}(\xi, \theta) &=& A_0(\xi) + \theta\,\big[\bar C^m\,\partial_m\, A_0  
- (\partial_0 \, \bar C^1 - \partial_1 \,\bar C^0) \, A_2 \nonumber\\
&-& (\partial_0 \, \bar C^1 + \partial_1 \, \bar C^0)\,  A_1\big]  \equiv A_0(\xi) + \theta\,\big[s_{ab}\,A_0(\xi)\big], \nonumber\\
A_1(\xi) \quad \rightarrow \quad {\cal A}^{(c)}_{1(ab)}(\xi, \theta) &=& A_1(\xi) + \theta\,\big[\bar C^m\,\partial_m \, A_1  
- (\partial_1 \,\bar C^0 + \partial_0 \, \bar C^1)\, A_0 \nonumber\\
&-& (\partial_0 \, \bar C^0 - \partial_1 \, \bar C^1) \, A_2\big]  \equiv A_1(\xi) + \theta\,\big[s_{ab}\,A_1(\xi)\big], \nonumber\\
A_2(\xi) \quad \rightarrow \quad {\cal A}^{(c)}_{2(ab)}(\xi, \theta) &=& A_2(\xi) + \theta\,\big[\bar C^m\,\partial_m \, A_2 - 
(\partial_1 \, \bar C^0 - \partial_0 \, \bar C^1) A_0 \nonumber\\
&-& (\partial_0 \, \bar C^0 - \partial_1 \, \bar C^1) A_1\big]  \equiv A_2(\xi) + \theta\,\big[s_{ab}\,A_2(\xi)\big]. 
\end{eqnarray}
In the above, the coefficients of $\theta$ are nothing but the anti-BRST symmetry transformations for the component fields
$A_0, \, A_1$ and $A_2$ [cf. Eq. (8)]. We point out that the subscripts $(b)$ and $(ab)$ in equation (60) and (61) have their straightforward meaning 
as we have established earlier. 
We lay emphasis on the fact that the super expansions in (60) and (61) are very
crucial and important as will be clear in the next section where we shall discuss the symmetry invariances.

\section{Invariance of the Lagrangian Densities: ACSA}

In this section, we capture the (anti-)BRST invariance of the Lagrangian densities (1) and (2) in terms of the (anti-)chiral superfields 
that have been obtained after the imposition of the (anti-)BRST invariant restrictions. In this connection, 
it is worth pointing out that we have already computed the BRST invariance of
${\cal L}_b$ {\it and} anti-BRST invariance of ${\cal L}_{ab}$ in the {\it ordinary} space in our earlier work [17]. To be precise, the action integrals
$S_1 = \int d^2\,\xi\,{\cal L}_b$ and $S_2 = \int d^2\,\xi\,{\cal L}_{ab}$ remain invariant under the continuous, infinitesimal and
nilpotent transformations in (7) and (6). In this connection, first of all, we note that the following are true for the 
{\it classical} Lagrangian density $({\cal L}_0)$, namely;
\begin{eqnarray}
s_b\,{\cal L}_0 = \partial_a\,\big[C^a\,{\cal L}_0\big], \qquad \qquad  s_{ab}\,{\cal L}_0 = \partial_a\,\big[\bar C^a\,{\cal L}_0\big],
\end{eqnarray}
and the {\it total} Lagrangian densities ${\cal L}_b$ and ${\cal L}_{ab}$ transform as [17]:
\begin{eqnarray}
s_b\, {\cal L}_{b} &=& \partial_a \, \Bigl[C^a \big({\cal L}_{0} + B_0\,A_0 + B_1\,A_1\big) + i\, \bar C_1 C^b\,\partial_b \, (C^a\,A_1) 
+ i\, \bar C_1 \, C^a (\partial_0 \, C^1 + \partial_1 \, C^0) \, A_0 \nonumber\\
&+& i\, \bar C_0 C^b\,\partial_b \, (C^a\,A_0) + i\,\bar C_0 \, C^a \,(\partial_0 \, C^1 + \partial_1 \, C^0) \, A_1 
+ i\,\bar C_0 \, C^a \,(\partial_0 \, C^1 - \partial_1 \, C^0) \, A_2 \nonumber\\ 
&+& i\,\bar C_1 \, C^a \,(\partial_0 \, C^0 - \partial_1 \, C^1) \, A_2\Bigr].  
\end{eqnarray}
\begin{eqnarray}
s_{ab}\, {\cal L}_{ab} &=& \partial_a \, \Bigl[\bar C^a \big({\cal L}_{0} - \bar B_0\,A_0 - \bar B_1\,A_1\big) 
- i\,  C_1\,\bar C^b\,\partial_b \, (\bar C^a\,A_1) - i\, C_1 \, \bar C^a (\partial_0 \, \bar C^1 + \partial_1 \, \bar C^0) \, A_0 \nonumber\\
&-& i\, C_0 \,\bar C^b\,\partial_b \, (\bar C^a\,A_0)
- i\,C_0 \, \bar C^a \,(\partial_0 \, \bar C^1 + \partial_1 \, \bar C^0) \, A_1 
- i\, C_0 \, \bar C^a \,(\partial_0 \, \bar C^1 - \partial_1 \, \bar C^0) \, A_2 \nonumber\\
&-& i\, C_1 \, \bar C^a \,(\partial_0 \, \bar C^0 - \partial_1 \, \bar C^1) \, A_2\Bigr]. 
\end{eqnarray}
The above observations demonstrate that $s_b\,S_1 = 0$ and $s_{ab}\,S_2 = 0$ for the physical fields of the (anti-)BRST
invariant theories which vanish-off [16] at $\sigma = 0,\, \pi$ and $\tau \rightarrow  \pm\,\infty$ due to Gauss's divergence theorem.
We mention, in passing, that $s_b\,S_0 = 0$ and $s_{ab}\,S_0 = 0$ (where $S_0 = \int d^2\,\xi\,{\cal L}_0$) due to the
(anti-)BRST transformations for ${\cal L}_0$ in (62).

First of all, we capture the (anti-)BRST invariance of the action integral $S_0 = \int d^2\,\xi\,{\cal L}_0$ within the
realm of ACSA. In this regard, we note the following (anti-)chiral generalizations of ${\cal L}_0$ to its {\it counterpart}
super Lagrangians (i.e. ${\cal L}_0 \rightarrow \tilde{\cal L}_0$) on the $(2, 1)$-dimensional (anti-) chiral super
submanifolds, namely;
\begin{eqnarray}
{\cal L}_0 \quad \rightarrow \quad \tilde{\cal L}_0^{(ac)}(\xi, \bar\theta) &=& - \frac{1}{2k}\,{\tilde G}^{mn(ac)}_{(b)}(\xi, \bar\theta)\,
\partial_m\,{\tilde X}^{\mu(h, ac)}(\xi, \bar\theta)\,\partial_n\,{\tilde X}_{\mu}^{(h, ac)}(\xi, \bar\theta) \nonumber\\ 
&+& {\cal E}^{(ac)}_{(b)}(\xi, \bar\theta)
\,\big[\mbox{det}\,\tilde G^{(ac)}_{(b)}(\xi, \bar\theta) + 1\big], \nonumber\\
{\cal L}_0 \quad \rightarrow \quad \tilde{\cal L}_0^{(c)}(\xi, \theta) &=& - \frac{1}{2k}\,{\tilde G}^{mn(c)}_{(ab)}(\xi, \theta)\,
\partial_m\,{\tilde X}^{\mu(h, c)}(\xi, \theta)\,\partial_n\,{\tilde X}_{\mu}^{(h, c)}(\xi, \theta) \nonumber\\ 
&+& {\cal E}^{(c)}_{(ab)}(\xi, \theta)
\,\big[\mbox{det}\,\tilde G^{(c)}_{(ab)}(\xi, \theta) + 1\big], 
\end{eqnarray}
where the super Lagrangian densities (on the l.h.s.) carry superscripts $(ac)$ and $(c)$ 
to denote that {\it these} have been defined on the (2, 1)-dimensional (anti-)chiral super submanifolds of the (2, 2)-dimensional
{\it general} supermanifold (that has been chosen for our discussion). The superfields with subscripts $(b)$ and $(ab)$
as well as with superscripts $(ac)$, $(c)$, $(h,\,c)$ and $(h,\,ac)$ have already been explained in our previous and present sections. The 
equation (62) can be captured in the superspace (where ACSA plays an important role). The 
mappings: $s_b\leftrightarrow \partial_{\bar \theta}$, $s_{ab}\leftrightarrow \partial_{\theta}$ lead to the following
observations:
\begin{eqnarray}
\frac{\partial}{\partial\,\bar\theta}\,\tilde{\cal L}_0^{(ac)}(\xi, \bar\theta) = \partial_a\,\big[C^a\,{\cal L}_0\big] \equiv s_b\,{\cal L}_0,
\qquad \;\; \frac{\partial}{\partial\,\theta}\,\tilde{\cal L}_0^{(c)}(\xi, \theta) = \partial_a\,\big[\bar C^a\,{\cal L}_0\big] \equiv s_{ab}\,{\cal L}_0. 
\end{eqnarray}
Thus, the (anti-)BRST symmetry invariances of ${\cal L}_0$ have been expressed in the language of ACSA to BRST formalism.
We have performed {\it this} exercise {\it separately} because, on its own, the {\it original} classical Lagrangian density ${\cal L}_0$ transforms 
to the {\it total} derivatives [cf. Eq. (62)] under the (anti-)BRST symmetry transformations.

We would like to express the symmetry transformations (63) and (64) in the realm of ACSA where the super expansions in
(27), (32), (33), (40), (44), (53) and (60) will be playing decisive roles for the BRST invariance [cf. Eq. (63)].
On the other hand, the super expansions (27), (57), (59) and (61) will be very {\it useful} in capturing the
anti-BRST invariance [cf. Eq. (64)].  With these inputs at our disposal, we set out to capture the BRST invariance in
terms of $\partial_{\bar \theta}$ and ${\tilde{\cal L}^{ac}_b}(\xi,\,\bar\theta)$. Here the {\it latter} is given 
in the language of the {\it anti-chiral} superfields that have been derived after the imposition of the BRST-invariant
restrictions. These anti-chiral superfields might {\it also} be the limiting cases of the full super expansions that have been 
derived in Sec. 3, namely;.
\begin{eqnarray}
\tilde{\cal L}_b^{(ac)}(\xi, \bar\theta) &=& \tilde{\cal L}_0^{(ac)}(\xi, \bar\theta) + B_0(\xi)\,{\cal A}_{0(b)}^{(ac)}(\xi, \bar\theta) 
+ B_1(\xi)\,{\cal A}_{1(b)}^{(ac)}(\xi, \bar\theta) \nonumber\\
&-& i\,\Big[{\bar F}_{1(b)}^{(ac)}(\xi, \bar\theta)\,
\Big\{\partial_0\,{F}^{1(ac)}_{(b)}(\xi, \bar\theta) + \partial_1\,{F}^{0(ac)}_{(b)}(\xi, \bar\theta)\Big\} \nonumber\\
&+& {\bar F}^{(ac)}_{0(b)}(\xi, \bar\theta)\,\Big\{\partial_a\,{F}^{a(ac)}_{(b)}(\xi, \bar\theta)\Big\} - {F}^{a(ac)}_{(b)}(\xi, \bar\theta)\,
\Big\{\partial_a\,{\bar F}^{(ac)}_{0(b)}(\xi, \bar\theta)\Big\}\Big]\,{\cal A}_{0(b)}^{(ac)}(\xi, \bar\theta) \nonumber\\
&-& i\,\Big[{\bar F}_{0(b)}^{(ac)}(\xi, \bar\theta)\,
\Big\{\partial_0\,{F}^{1(ac)}_{(b)}(\xi, \bar\theta) + \partial_1\,{F}^{0(ac)}_{(b)}(\xi, \bar\theta)\Big\} \nonumber\\
&-& {F}^{a(ac)}_{(b)}(\xi, \bar\theta)\,\Big\{\partial_a\,{\bar F}^{(ac)}_{1(b)}(\xi, \bar\theta)\Big\} + {\bar F}^{(ac)}_{1(b)}(\xi, \bar\theta)\,
\Big\{\partial_a\,{F}^{a(ac)}_{(b)}(\xi, \bar\theta)\Big\}\Big]\,{\cal A}_{1(b)}^{(ac)}(\xi, \bar\theta) \nonumber\\
&-& i\,\Big[{\bar F}_{1(b)}^{(ac)}(\xi, \bar\theta)\,
\Big\{\partial_0\,{F}^{0(ac)}_{(b)}(\xi, \bar\theta) - \partial_1\,{F}^{1(ac)}_{(b)}(\xi, \bar\theta)\Big\} \nonumber\\
&+& {\bar F}_{0(b)}^{(ac)}(\xi, \bar\theta)\,\Big\{\partial_0\,{F}^{1(ac)}_{(b)}(\xi, \bar\theta) 
- \partial_1\,{F}^{0(ac)}_{(b)}(\xi, \bar\theta)\Big\}\Big]\,{\cal A}_{2(b)}^{(ac)}(\xi, \bar\theta),
\end{eqnarray}
where we have taken the {\it ordinary} fields $B_0 (\xi)$ and $B_1(\xi)$ because we know that $B^m(\xi)\rightarrow 
{\cal B}^{m}_{(b)} (\xi,\,\bar\theta) = B^m (\xi)$ due to the BRST invariance $[s_b\, B^m (\xi) = 0]$ of $B^m(\xi)]$. Ultimately,
it turns out that we obtain the following due to operation of $\partial_{\bar\theta}$ on $\tilde{\cal L}_b^{(ac)}(\xi, \bar\theta)$:
\begin{eqnarray}
\frac{\partial}{\partial\,\bar\theta}\,\tilde{\cal L}_b^{(ac)}(\xi, \bar\theta) &=& \partial_a \, \Bigl[C^a \big({\cal L}_{0} + B_0\,A_0 
+ B_1\,A_1\big) + i\, \bar C_1 C^b\,\partial_b \, (C^a\,A_1) \nonumber\\
&+& i\, \bar C_1 \, C^a (\partial_0 \, C^1 + \partial_1 \, C^0) \, A_0 
+ i\, \bar C_0 C^b\,\partial_b \, (C^a\,A_0) \nonumber\\ 
&+& i\,\bar C_0 \, C^a \,(\partial_0 \, C^1 + \partial_1 \, C^0) \, A_1 
+ i\,\bar C_0 \, C^a \,(\partial_0 \, C^1 - \partial_1 \, C^0) \, A_2  \nonumber\\
&+& i\,\bar C_1 \, C^a \,(\partial_0 \, C^0 - \partial_1 \, C^1) \, A_2\Bigr] \equiv s_b\,{\cal L}_b.  
\end{eqnarray}
It is evident that the {\it above} equation captures the BRST invariance of the Lagrangian density ${\cal L}_b$ in the 
{\it superspace} (as is clear from our observation on the r.h.s.).

We can repeat the {\it same} exercise for the anti-BRST invariance. For this purpose, first of all, we generalize ${\cal L}_{ab}$
to its counterpart {\it chiral} super Lagrangian density on the (2, 1)-dimensional {\it chiral} super submanifold as  
\begin{eqnarray}
\tilde{\cal L}_{ab}^{(c)}(\xi, \theta) &=& \tilde{\cal L}_0^{(c)}(\xi, \theta) - \bar B_0(\xi)\,{\cal A}_{0(ab)}^{(c)}(\xi, \theta) 
- \bar B_1(\xi)\,{\cal A}_{1(ab)}^{(c)}(\xi, \theta) \nonumber\\
&+& i\,\Big[{F}_{1(ab)}^{(c)}(\xi, \theta)\,
\Big\{\partial_0\,{\bar F}^{1(c)}_{(ab)}(\xi, \theta) + \partial_1\,{\bar F}^{0(c)}_{(ab)}(\xi, \theta)\Big\} \nonumber\\
&+& {F}^{(c)}_{0(ab)}(\xi, \theta)\,\Big\{\partial_a\,{\bar F}^{a(c)}_{(ab)}(\xi, \theta)\Big\} + 
\Big\{\partial_a\,{F}^{(c)}_{0(ab)}(\xi, \theta)\Big\}\,{\bar F}^{a(c)}_{(ab)}(\xi, \theta)\Big]\,{\cal A}_{0(ab)}^{(c)}(\xi, \theta) \nonumber\\
&+& i\,\Big[{F}_{0(ab)}^{(c)}(\xi, \theta)\,
\Big\{\partial_0\,{\bar F}^{1(c)}_{(ab)}(\xi, \theta) + \partial_1\,{\bar F}^{0(c)}_{(ab)}(\xi, \theta)\Big\} \nonumber\\
&+& {F}^{(c)}_{1(ab)}(\xi, \theta)\,
\Big\{\partial_a\,{\bar F}^{a(c)}_{(ab)}(\xi, \theta)\Big\} + \Big\{\partial_a\,{F}^{(c)}_{1(ab)}
(\xi, \bar\theta)\Big\}\,{\bar F}^{a(c)}_{(ab)}(\xi, \theta)\Big]\,{\cal A}_{1(ab)}^{(c)}(\xi, \theta) \nonumber\\
&+& i\,\Big[{F}_{1(ab)}^{(c)}(\xi, \theta)\,
\Big\{\partial_0\,{\bar F}^{0(c)}_{(ab)}(\xi, \theta) - \partial_1\,{\bar F}^{1(c)}_{(ab)}(\xi, \theta)\Big\} \nonumber\\
&+& {F}_{0(ab)}^{(c)}(\xi, \theta)\,\Big\{\partial_0\,{\bar F}^{1(c)}_{(ab)}(\xi, \theta) 
- \partial_1\,{\bar F}^{0(c)}_{(ab)}(\xi, \theta)\Big\}\Big]\,{\cal A}_{2(ab)}^{(c)}(\xi, \theta),
\end{eqnarray}
where the ordinary fields ${\bar B}_0(\xi)$ and ${\bar B}_1(\xi)$ are present in the above {\it super} Lagrangian density because
$s_{ab}\,{\bar B}^a = 0$ which implies that ${\bar B}^a(\xi)  \rightarrow  \bar{\cal B}_{(ab)}^{a(c)}(\xi, \theta) = {\bar B}^a(\xi) + 
\theta\,(0) \equiv {\bar B}^a(\xi)$. In other words, there is {\it no} chiral $\theta$-dependence on the r.h.s. of the super expansion of the
 superfield $\bar{\cal B}_{(ab)}^{a(c)}(\xi, \theta)$.
The {\it rest} of the notations for the {\it chiral} superfields have already been explained earlier. At this juncture, in 
view of the mapping: $s_{ab} \leftrightarrow \partial_{\theta}$, we can capture the anti-BRST invariance (64) by applying
a derivative $\partial_{\theta}$ on (69) which yields the following:
\begin{eqnarray}
\frac{\partial}{\partial\,\theta}\,\tilde{\cal L}_{ab}^{(c)}(\xi, \theta) &=& \partial_a \, \Bigl[\bar C^a \big({\cal L}_{0} - \bar B_0\,A_0 
- \bar B_1\,A_1\big) - i\,  C_1\,\bar C^b\,\partial_b \, (\bar C^a\,A_1) \nonumber\\
&-& i\, C_1 \, \bar C^a (\partial_0 \, \bar C^1 + \partial_1 \, \bar C^0) \, A_0 - i\, C_0 \,\bar C^b\,\partial_b \, (\bar C^a\,A_0) \nonumber\\
&-& i\,C_0 \, \bar C^a \,(\partial_0 \, \bar C^1 + \partial_1 \, \bar C^0) \, A_1 
- i\, C_0 \, \bar C^a \,(\partial_0 \, \bar C^1 - \partial_1 \, \bar C^0) \, A_2 \nonumber\\
&-& i\, C_1 \, \bar C^a \,(\partial_0 \, \bar C^0 - \partial_1 \, \bar C^1) \, A_2\Bigr] \equiv s_{ab}\,{\cal L}_{ab}.   
\end{eqnarray}
Hence we have captured the anti-BRST symmetry invariance (64) in the language of ACSA to BRST formalism [as is evident from 
the r.h.s. of (70)].

We close this section with the following remark. We can capture the basic ideas behind the derivations of ${\cal L}_b$ and ${\cal L}_{ab}$
which have been explained in Eq. (5). In view of the mappings: $s_b \leftrightarrow \partial_{\bar\theta},\, s_{ab} \leftrightarrow \partial_{\theta}$,
we can express the {\it super} (anti-)BRST invariant Lagrangian densities corresponding to the {\it ordinary} Lagrangian densities [cf. Eq. (5)] as
\begin{eqnarray}  
\tilde{\cal L}_{ab}^{(c)}(\xi, \theta) &=& {\cal L}_0^{(c)}(\xi, \theta) + \frac{\partial}{\partial\,\theta}\,
\Big[i\,F^{(c)}_{0(ab)}(\xi, \theta)\,{\cal A}^{(c)}_{0(ab)}
(\xi, \theta) + i\,F^{(c)}_{1(ab)}(\xi, \theta)\,{\cal A}^{(c)}_{1(ab)}(\xi, \theta)\Big], \nonumber\\
\tilde{\cal L}_{b}^{(ac)}(\xi, \bar\theta) &=& {\cal L}_0^{(ac)}(\xi, \bar\theta) + \frac{\partial}{\partial\,\bar\theta}\,
\Big[- i\,\bar F^{(ac)}_{0(b)}(\xi, \bar\theta)\,{\cal A}^{(ac)}_{0(b)}
(\xi, \bar\theta) - i\,\bar F^{(ac)}_{1(b)}(\xi, \bar\theta)\,{\cal A}^{(ac)}_{1(b)}(\xi, \bar\theta)\Big],
\end{eqnarray}
where {\it all} the symbols have been explained in our earlier discussion. It is crystal clear, from the above expression, that
the (anti-)BRST invariance of the action integrals $S_1 = \int d^2\,\xi\,{\cal L}_b$ and $S_2 = \int d^2\,\xi\,{\cal L}_{ab}$
can be captured in the terminology of ACSA to BRST formalism because $s_b\,S_1$ and $s_{ab}\,S_2$ will be {\it zero} in the {\it ordinary}
space. Furthermore, we note that $\partial_{\theta}\,\tilde{\cal L}_{ab}^{(c)}(\xi, \theta)$ and $\partial_{\bar\theta}\,
\tilde{\cal L}_{b}^{(ac)}(\xi, \bar\theta)$ will always produce the {\it total} derivatives in the {\it ordinary} space thereby rendering 
the action integrals (i.e. $S_1$ and  $S_2$) equal to zero [cf. Eq. (71)]. To be precise, the nilpotency $(\partial_{\bar\theta}^2 = 0,\,
\partial_{\theta}^2 = 0)$ property of the translational generators $(\partial_{\theta},\, \partial_{\bar\theta})$ will ensure that
$\partial_{\theta}\,\tilde{\cal L}_{ab}^{(c)}(\xi, \theta)$ and $\partial_{\bar\theta}\,\tilde{\cal L}_{b}^{(ac)}(\xi, \bar\theta)$ 
will be {\it always} the total derivatives in the {\it ordinary} space. Hence, we are able to capture the symmetry invariance(s) 
of the action integrals (corresponding to the Lagrangian densities ${\cal L}_b $  and ${\cal L}_{ab}$) using ACSA.

\section{Conclusions}

In our present endeavor, we have exploited the theoretical potential of MBTSA and ACSA to derive {\it all} 
the (anti-)BRST symmetry transformations for the 2D diffeomorphism symmetry invariant model of a bosonic string theory. These 
symmetry transformations [$s_{(a)b}$] are {\it proper} because they are off-shell nilpotent $[s_{(a)b}^2 = 0]$ of order two {\it and} absolutely 
anticommuting (i.e. $s_b\,s_{ab} + s_{ab}\,s_b = 0$) in nature [cf. Eqs. (10),(7),(6)]. The {\it latter} property of 
the (anti-)BRST symmetry transformations [$s_{(a)b}$] is satisfied if and only if we invoke the sanctity of the 
CF-type restrictions: $ B^a + \bar B^a + i\, (C^m \, \partial_m \, \bar C^a +
 \bar C^m \, \partial_m \, C^a) = 0 $ (with $a, m = 0,1$) which define a {\it submanifold} in the {\it quantum} Hilbert 
space of fields {\it where} the Nakanishi-Lautrup type auxiliary fields as well as the (anti-)ghost fields are present algebraically in 
a specific manner [cf. Eq. (24)]. These restrictions are {\it physical} in some sense because they are (anti-)BRST symmetry invariant
[cf. Eqs. (7),(6)] on the above {\it submanifold}. Hence, their imposition on our BRST-quantized theory is logical.

By applying the theoretical strength of MBTSA, we have been able to derive, in one stroke, the (anti-)BRST 
symmetry transformations {\it together} for the Lorentz pure {\it scalar} fields [e.g. $ X^\mu (\xi),\, (\mbox{det} \,\tilde g) $] 
and the 2D version of the {\it universal} CF-type restrictions: $ B^a + \bar B^a + i\, (C^m \, \partial_m \, \bar C^a +
 \bar C^m \, \partial_m \, C^a) = 0 $. These 2D restrictions are the limiting case of the D-dimensional 
diffeomorphism invarant theory where the superfield approach (developed by us [18, 13]) leads 
to the existence of the D-dimensional CF-type restrictions: $ B_\mu + \bar B_\mu + i\, (C^\rho \, \partial_\rho \, \bar C_\mu +
 \bar C^\rho \, \partial_\rho \, C_\mu) = 0 $ (with $\mu = 0,\,1,\,2...D-1$) where the {\it fermionic}
(anti-)ghost fields $(\bar C_\mu)C_\mu$ correspond to the D-dimensional infinitesimal and
continuous diffeomorphism symmetry transformations: $x_\mu\rightarrow x_\mu^{'} = x_\mu - \epsilon_{\mu} (x)$.
In {\it these} infinitesimal transformations, the parameters $\epsilon_\mu (x) $ are the diffeomorphism transformation parameters.
The symbols $(\bar B_\mu)B_\mu$ are nothing but the Nakanishi-Lautrup type auxiliary fields in the D-dimensional theory.
The existence of the D-dimensional CF-type restrictions: $B^{\mu} + \bar B^{\mu} + i\, (C^{\rho} \, \partial_{\rho} \, \bar C^{\mu} +
 \bar C^{\rho} \, \partial_{\rho} \, C^{\mu}) = 0$ are {\it universal} and, so far, their presence have been shown
{\it explicitly} in the cases of 2D and 1D diffeomorphism invariant theories (see, e.g. [17, 14] for details).

Within the ambit of MBTSA, it becomes evident that we have to take, at least, the helps of the {\it (anti-)chiral} superfield
expansions [cf. Eq. (27)] so that we can obtain $s_b\,\bar C_\mu = i\,B_\mu$ and $s_{ab}\,C_\mu = i\,\bar B_\mu$ for the
D-dimensional diffeomorphism invariant theory (see e.g. [13, 18] for details) in addition to the validity of off-shell 
{\it nilpotency} property so that we can obtain: $s_b\,C_\mu = C^\rho\,\partial_\rho\,C_\mu$ and  
$s_{ab}\,\bar C_\mu = \bar C^\rho\,\partial_\rho\,\bar C_\mu$. The above {\it two} inputs are essential for the completeness 
of MBTSA. Hence, we have exploited the theoretical potential of the ACSA to BRST formalism (see, e. g. [15]) so that {\it both} the above 
{\it inputs} can be taken care of. As a 
consequence, it becomes important to blend {\it together} the ideas from the MBTSA and ACSA so that we can derive {\it all} the (anti-)BRST symmetry
transformations for the {\it all} the fields of a diffeomorphism invariant theory {\it along with} the derivation  of appropriate 
(anti-)BRST invariant CF-type restrictions. This is what we have {\it precisely} done in our present investigation.
Our earlier works (see, e.g. [14] and references theirin) on the 1D diffeomorphism invariant models of the relativistic and non-relativistic particles (of
SUSY and non-SUSY varieties) have {\it also} exploited the ideas behind MBTSA and ACSA {\it together} to obtain the 1D version 
$[B + \bar B + i\, (\bar C \, \dot C - \dot{\bar C}\, C)  = 0]$ of the {\it universal} D-dimensional CF-type restrictions
that have been derived and  thoroughly discussed in [18, 13].

In our earlier work [17] on our {\it present} bosonic string, we have computed the expressions for the BRST and anti-BRST
charges in the {\it flat} space. In the paper by Kato and Ogawa [16], the nilpotency of BRST charge has been proven
to demonstrate that {\it quantum} version of the theory is valid {\it only} when $D = 26$ and $\alpha_0 = 1$. It will be
very nice future endeavor for us to take the expression for the anti-BRST charge and plug in the {\it normal} mode expansions
of the fields (with creation and annihilation operators in it) so that the {\it quantum} version of {\it it} can be obtained.
With appropriate {\it boundary conditions} on the {\it target space} coordinate fields and (anti-)ghost fields, it will be
challenging to derive $D = 26$ and $\alpha_0 = 1$ from the requirement of the {\it nilpotency} of the {\it anti-BRST} charge
in the flat limit. We are presently involved with {\it this} problem and our results/observations will be reported elsewhere.

As pointed out earlier, our present 2D diffeomorphism invariant theory is {\it different} from our earlier works on the 1D diffeomorphism
(i.e. reparameterization) invariant theories  (see, e.g. [14] and references therein) in the sense that the {\it latter} theoretical models
have the gauge symmetry transformations, too, which are equivalent to the reparameterization (i.e. 1D diffeomorphism)
symmetry transformations in the {\it specific} limits (see, e.g. [19, 14] for details). It is worth emphasizing that the gauge symmetry 
transformations (generated by the first-class constraints) have been exploited for the  BRST quantization in [19] in the cases of the
1D diffeomorphism (i.e. reparameterization) invariant models. The {\it latter} models are nothing but the non-SUSY scalar relativistic and SUSY spinning 
relativistic particles. We lay emphasis on the fact that the reparameterization symmetry  transformations of 
{\it these} models have been left untouched in [19] as far as the BRST quantization scheme is concerned.
We have taken this challenge in our earlier works (see, e.g. [14] and references therein) for the BRST quantization of these models.

\vskip 0.4cm

\noindent
{\bf Acknowledgments} \vskip 0.2cm

\noindent
Two of us (AT and AKR) would like to express their deep sense of gratefulness towards Banaras Hindu University (BHU) for {\it its}
financial support through the BHU-fellowship. Prof G. Rajasekaran is one of the very prominent {\it mentors} of our group and {\it all} 
three of us would like to dedicate  {\it this} work, very humbly and respectfully, to {\it him} on the occasion of his $85^{th}$ birth-anniversary which 
was celebrated on $22^{nd}$ February 2021.

\vskip 0.3cm

\noindent
{\bf Data Availability}\\

\noindent
No data were used to support this study.\\

\noindent
{\bf Conflicts of Interest}\\

\noindent
The authors declare that there are no conflicts of interest.

\end{document}